\documentclass[fleqn,usenatbib]{mnras}

\usepackage{newtxtext,newtxmath}
\usepackage[normalem]{ulem}
\usepackage{threeparttable}
\usepackage[T1]{fontenc}
\usepackage{tabularx}
\usepackage{longtable}
\usepackage{booktabs}
\usepackage{siunitx}
\usepackage{xcolor}

\DeclareRobustCommand{\VAN}[3]{#2}
\let\VANthebibliography\thebibliography
\def\thebibliography{\DeclareRobustCommand{\VAN}[3]{##3}\VANthebibliography}
\newcommand{\muHz}{\mbox{$\mu$}Hz}

\newcommand{\numax}{\mbox{$\nu_{\rm max}$}}
\newcommand{\dnu}{\mbox{$\Delta\nu$}}

\newcommand{\numaxsun}{\mbox{$\nu_{\rm max,\odot}$}}
\newcommand{\dnusun}{\mbox{$\Delta\nu_{\odot}$}}

\newcommand{\Teff}{\mbox{$T_{\rm eff}$}}
\newcommand{\Teffsun}{\mbox{$T_{\rm eff, \odot}$}}

\newcommand{\fnumax}{\mbox{$f_{\nu_{\rm max}}$}}
\newcommand{\fdnu}{\mbox{$f_{\dnu}$}}

\newcommand{\msun}{\mbox{$\rm M_{\odot}$}}
\newcommand{\rsun}{\mbox{$\rm R_{\odot}$}}
\newcommand{\gsun}{\mbox{$\rm g_{\odot}$}}
\newcommand{\logg}{\mbox{$\log g$}}

\newcommand{\gbp}{\mbox{$\rm G_{BP}$}}
\newcommand{\grp}{\mbox{$\rm G_{RP}$}}

\newcommand{\corot}{{CoRoT\/}}
\newcommand{\kepler}{{\em Kepler\/}}

\newcommand{\echelle}{{\'e}chelle}

\newcommand{\fullsample}{{72,647}}
\newcommand{\yes}{{11,349}}
\newcommand{\maybe}{{7804}}

\newcommand{\visoc}{{19,151}}
\newcommand{\atlok}{{18,430}}

\newcommand{\nblend}{{813}}
\newcommand{\fullosc}{17,617}

\newcommand{\gooddnu}{{10,298}}
\newcommand{\gooddnuRGB}{{7469}}
\newcommand{\gooddnuRC}{{2723}}

\newcommand{\xgbgood}{{3455}}

\newcommand{\goldall}{{5226}}
\newcommand{\goldRGB}{{4061}}
\newcommand{\goldRC}{{1165}}

\newcommand{\pysydfail}{{169}}

\newcommand{\fullsamplesouth}{{9667}}
\newcommand{\fullsamplenorth}{{7950}}

\newcommand{\lastsector}{{87}}

\newcommand{\fullnumaxprec}{{1.5}}
\newcommand{\fulldnuprec}{{1.0}}
\newcommand{\goldnumaxprec}{{1.3}}
\newcommand{\golddnuprec}{{0.6}}

\newcommand{\orcid}[1]{\href{https://orcid.org/#1}{\textsuperscript{\includegraphics[width=10pt]{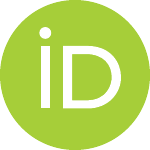}}}}

\usepackage{graphicx}	
\usepackage{amsmath}	

\title[Oscillating Red Giants in TESS CVZs]{ Global asteroseismology of 19,000 red giants in the TESS Continuous Viewing Zones}

\author[Sreenivas et al.]{
K. R. Sreenivas \orcid{0000-0003-1179-2069}$^{1}$\thanks{E-mail: skal9597@uni.sydney.edu.au},
Timothy R. Bedding\orcid{0000-0001-5222-4661}$^{1}$,
Daniel Huber\orcid{0000-0001-8832-4488}$^{1,2}$,
Dennis Stello\orcid{0000-0002-4879-3519}$^{3}$,
Marc Hon\orcid{0000-0003-2400-6960}$^{4,5}$,
\newauthor{}
Claudia Reyes\orcid{0000-0001-9632-2706}$^{6}$, 
Yaguang Li\orcid{0000-0003-3020-4437}$^{2}$,
Daniel Hey\orcid{0000-0003-3244-5357}$^{2}$
\\
$^1$Sydney Institute for Astronomy, School of Physics, University of Sydney, NSW 2006, Australia. \\
$^2$ Institute for Astronomy, University of Hawai`i, 2680 Woodlawn Drive, Honolulu, HI 96822, USA\\
$^3$School of Physics, University of New South Wales, Sydney, NSW 2052, Australia.\\
$^4$ Kavli Institute for Astrophysics and Space Research, Massachusetts Institute of Technology, 77 Massachusetts Avenue, Cambridge, MA 02139, USA\\
$^5$Department of Physics, National University of Singapore, 21 Lower Kent Ridge Road, Singapore, 119077\\
$^6$ Research School of Astronomy and Astrophysics, Australian National University, Canberra, Australian Capital Territory, Australia
}

\date{Accepted XXX. Received YYY; in original form ZZZ}

\pubyear{\the\year{}}

\begin{document}
\label{firstpage}
\pagerange{\pageref{firstpage}--\pageref{lastpage}}
\maketitle

\begin{abstract}
TESS (Transiting Exoplanet Survey Satellite) has produced long-term photometry for millions of stars across the sky. In this work, we present an asteroseismic catalogue of \visoc{} red giants in the TESS Continuous Viewing Zones using sectors 1--\lastsector{} (Years 1--7). We visually assessed the power spectra for oscillations, and then applied the computationally efficient \texttt{nuSYD} method to confirm reliability. We identified an increase of 80\% in the number of previously known oscillating red giants at a TESS magnitude $>$ 8. We determined the frequency of maximum power (\numax{}) and the large frequency separation (\dnu{}) using the \texttt{pySYD} pipeline, achieving typical precisions of \fullnumaxprec{}\,\% and \fulldnuprec{}\,\%, respectively. We classified the stars into Red Giant Branch (RGB) and Core Helium Burning (CHeB) classes using a Convolutional Neural Network. Using spectroscopic data for \gooddnu{} stars with reliable asteroseismic measurements, we have been able to measure stellar mass and radii with precisions of 7.5\% and 2.8\%, which is comparable to that from 4-yr \kepler{} data. 
A comparison of the seismic radii with Gaia radii shows excellent agreement. 
With three years of TESS data, the asteroseismic parameters are precise enough to identify the RGB bump and delineate the Zero Age Helium Burning edge. Combined with astrometric data, these parameters reveal established trends across the Galactic plane, providing a valuable set of uniformly determined asteroseismic parameters for Galactic Archaeology.
\end{abstract}




\section{Introduction} 
Asteroseismology, the study of stars through sound the waves propagating within their interiors, has significantly advanced our understanding of stellar structure and evolution. These standing waves, which result from the stochastic excitation and damping caused by stellar convection, provide valuable information about both the interiors and surface properties of stars. Observationally, these waves appear as distinct frequencies in Fourier space, forming a Gaussian-shaped power excess. This feature is characterized by two key parameters: the frequency of maximum power (\numax{}) and the large frequency separation (\dnu{}) \citep{ulrich1986,brown1991, kjb1995}. The frequency \numax{} is approximately proportional to the acoustic cutoff frequency, which is itself a function of the star's surface gravity and effective temperature. The large frequency separation, \dnu{}, reflects the average density of the star. The scaling relations connect these two frequencies to fundamental stellar parameters, allowing the determination of a star's mass, radius, and surface gravity \citep{stello_2008_wire, kallinger2010_corot}. These scaling relations have been extensively validated using radial velocity and photometric data, and they have proven to be a reliable method for determining precise stellar parameters in red giants.

The long-baseline photometry provided by space missions such as \corot{} and \kepler{} \citep{baglin2006, borucki2010} has revolutionized our understanding of red giants over the past decades. Ensemble studies of red giants \citep{chalinmigilo2013, jackewiez2021, noels2025} have demonstrated the power of precise photometry in characterizing stellar evolutionary states, internal structures, and rotation properties \citep{bedding2011, vrard2016, mosser2016, gehan2018}. \citet{yu18} produced a catalogue of 16,094 \kepler{} red giants and revealed a significant dependence of solar-like oscillation amplitudes on metallicity and mass. In conjunction with simulations, \citet{yaguang22} used \kepler{} red giant data to validate asteroseismic scaling relations and quantify the intrinsic scatter in these relations. Additionally, \kepler{} observations established a metallicity-dependent trend in the scaling relation for the frequency of maximum power (\fnumax{}) \citep{viani2017, yaguang2022, tanda2022, huber2024, mia2025}. These datasets were used to advance our understanding of red giants at a population level. They also supported the development of methods for asteroseismic analysis of large-scale datasets \citep{hon2018, honqlp, SREE24}. Because of the precise nature of asteroseismic measurements, these datasets enabled studies of the solar neighbourhood and the mapping of our Galaxy \citep{migilo2013, anders2017, k2gap, apokasc3}.

While previous space missions, such as \corot{} and \kepler{}, focused on specific regions of the sky, the Transiting Exoplanet Survey Satellite (TESS, \citealt{ricker2015}) instead observes nearly the entire sky, producing light curves with typical baselines of 27 days per sector. This broad sky coverage has enabled TESS to detect oscillations in thousands of red giants across the sky \citep{honqlp, m21, hat2023, z24, Grusnis_2025,Theodoridis_2026}. However, due to its small aperture size and short observation baselines for most fields, TESS is less effective at detecting oscillations in faint ($\rm T_{\rm mag}$$ > 10$) high-\numax{} red giants \citep{dennis22}. At the same time, the higher-amplitude low-\numax{} stars require longer baseline for localizing power excess. Fortunately, TESS's wide field of view and observing strategy create two Continuous Viewing Zones (CVZs), each covering 452.16 deg$^{2}$
around the ecliptic poles: the Southern CVZ (ecliptic latitude $<-78$ degrees, SCVZ) and the Northern CVZ (ecliptic latitude $>78$ degrees, NCVZ). The SCVZ benefits from a longer observing baseline, with light curves spanning 3 years, while the NCVZ has 2 years of continuous observations due to TESS's initial focus on the southern sky. 


In this work, we present a catalogue of oscillating red giants in TESS's CVZs, with asteroseismic masses and radii determined homogeneously. By focusing on this region, we aim to expand the sample of distant, low-luminosity red giants for use in future investigations.


\section{Data } \label{sec:style}
\subsection{Sample selection}

\begin{figure}
    \includegraphics[width=1\linewidth]{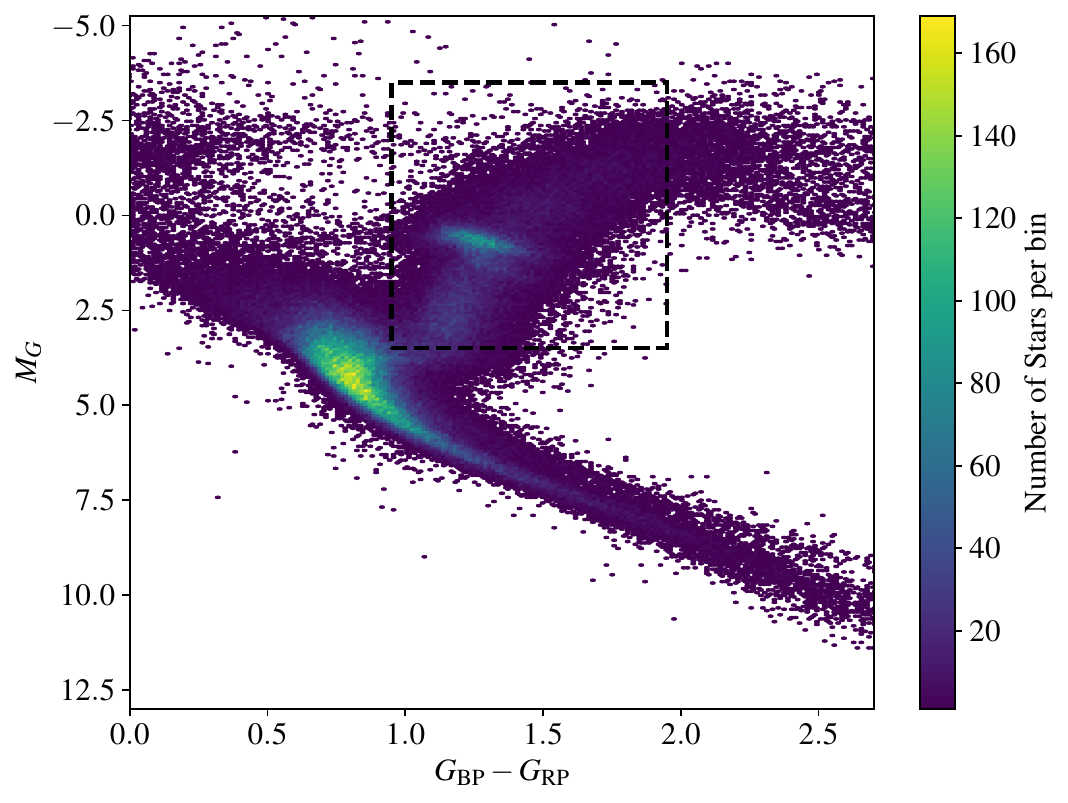}
    \caption{Colour-magnitude diagram using Gaia apparent magnitudes of all stars in TESS CVZ with TESS magnitude brighter than 13.5. The rectangle shows the \fullsample{} stars selected for analysis in this work. }
    \label{fig:cmd}
\end{figure}
\begin{figure}
    \centering
    \includegraphics[width=\linewidth]{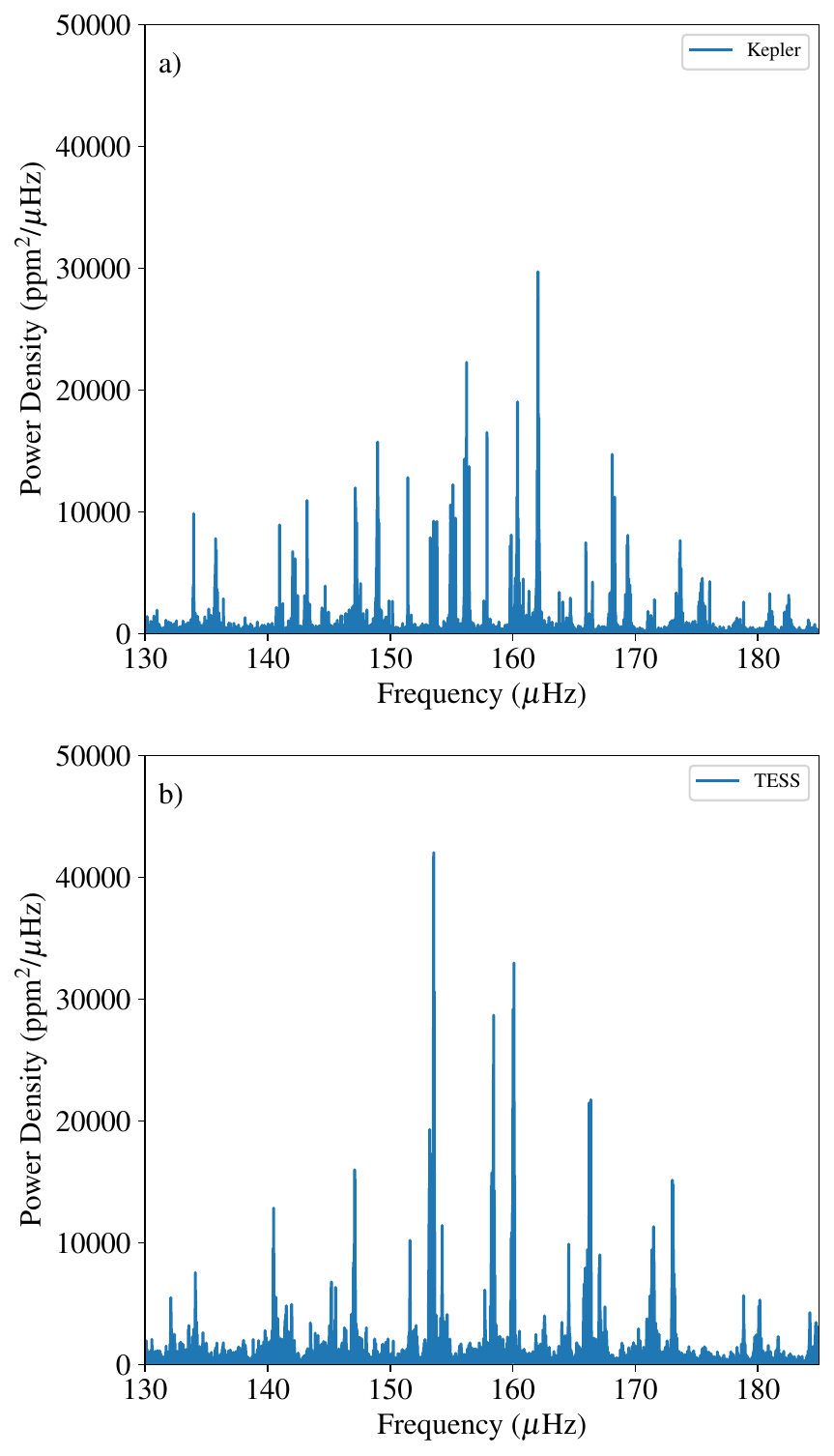}
    \caption{Solar like oscillations of a typical red giant. Panel a shows the power density spectra of an oscillating \kepler{} red giant (KIC\,9075872, $K_{p}$ = 11.89 ) and Panel b for an oscillating red giant (TIC\,237197414, $\rm T_{\rm mag}$ = 7.084) in the northern TESS CVZ.  }
    \label{fig:pds}
\end{figure}
From the TESS Input catalogue (TIC, version 8.2, \citealt{ticvp82}), we selected all stars with ecliptic latitude $<-78$\,deg (SCVZ) and $>78$\,deg (NCVZ). Due to the limited availability of light curves at fainter magnitudes and the difficulty in detecting oscillations, we restricted our sample to stars with TESS magnitude ($\rm T_{\rm mag}$) less than 13.5. We removed the duplicate entries by setting \texttt{q\_Gaia = 1} and \texttt{Disposition = NULL} columns in the TIC. The first criterion ensures that the stars have reliable astrometric information from Gaia \citep{gaiaval}, and the second ensures that the entries are unique \citep{ticvp82}. Figure \ref{fig:cmd} shows the colour-magnitude diagram using the absolute magnitude ($M_G$) and the dereddened colour indices (\gbp -- \grp) from Gaia DR2 \citep{gaiadr2} information from TIC. We selected all stars with $-3 < M_G < 4$ and $1 < \gbp - \grp < 2$, which left us with \fullsample{} possible red giants. As can be seen from the  fig. \ref{fig:cmd}, such a selection includes stars from the lower to the upper red giant branch and the red clump, on which this study will be focused.


\subsection{Light curves and power density spectra}
TESS observes the sky in 27-d sectors. During the first two years of the mission, the Full-Frame Images (FFIs) were produced at a cadence of 30 minutes, which was later reduced to 10 min (Year 3 onwards) and then to 200 seconds (Year 5 onwards). Two of the widely used light curve products from these FFIs are TESS-SPOC \citep{jenkins2016,caldwell2022} and QLP \citep{chelsea22, chelsea2}. Although QLP light curves are available across a greater number of sectors, we have found that TESS-SPOC light curves for red giants typically have lower noise levels \citep{hon22, sree25}.

We used data from TESS Sectors 1--\lastsector{} (Years 1--7). For each star in each sector, we used TESS-SPOC light curves whenever available and used QLP otherwise. 
We excluded cadences associated with any anomalies by setting the quality flag to 0\footnote{\url{https://archive.stsci.edu/missions/tess/doc/EXP-TESS-ARC-ICD-TM-0014.pdf}}. The light curves were high-pass filtered using a Gaussian kernel with a sigma of 5 days to remove low-frequency signals, and a 5-sigma clipping was applied to remove outliers. We calculated the power spectra using the Lomb–Scargle periodogram \citep{pressandrybicki}, implemented in the Python library \texttt{astropy}\footnote{\url{https://www.astropy.org/}}, and converted them to power density by multiplying by the effective observation time. We calculated the power spectrum up to 283.4\,\muHz{}, corresponding to the Nyquist frequency of the FFIs with a cadence of 30 min \citep{sree25}. Figure \ref{fig:pds} compares the power density spectra of an oscillating red giant in \kepler{} (top panel) and in the TESS CVZ (bottom panel). It can be seen that the oscillations in the TESS power spectrum has signal-to-noise ratiocomparable to that from 4-yr \kepler{} data.


\section{Detection and estimation of Global asteroseismic parameters} \label{sec:floats}
\subsection{Identifying oscillating red giants}

\begin{figure}
    \includegraphics[width=1.1\linewidth]{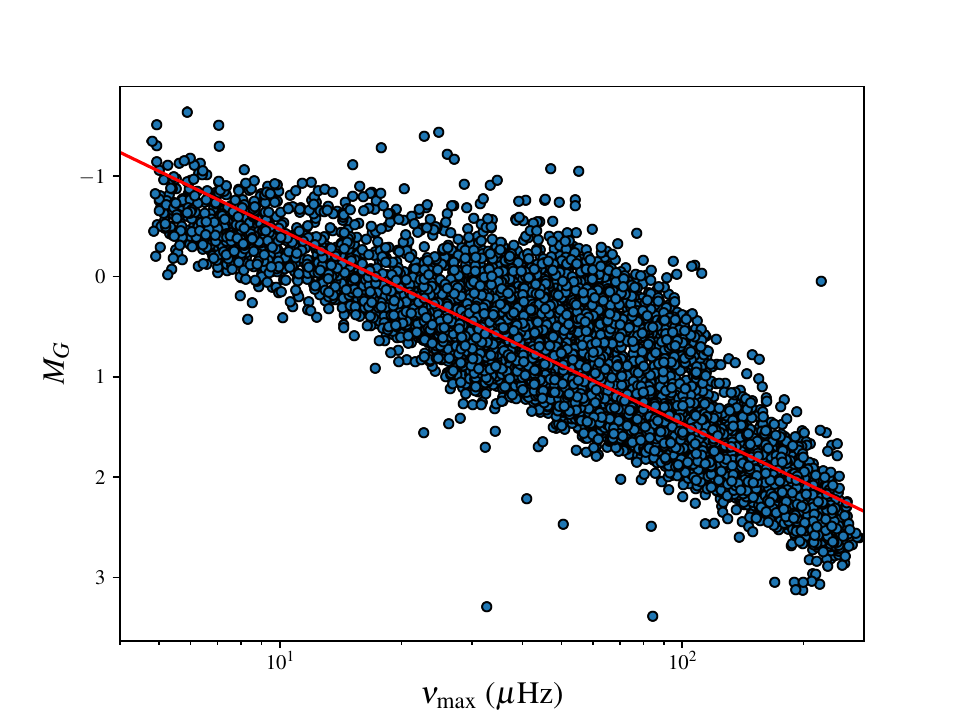}
    \caption{Gaia Absolute Magnitude ($ M_{G}$) vs Frequency of Maximum power (\numax{}) for all 16,094 \kepler{} stars in \citet{yu18}. The red line corresponds to the fit provided in eqn. \ref{eqn1}.}
    \label{fig:mgvsnumax}
\end{figure}
\begin{figure}
    \centering
    \includegraphics[width=\linewidth]{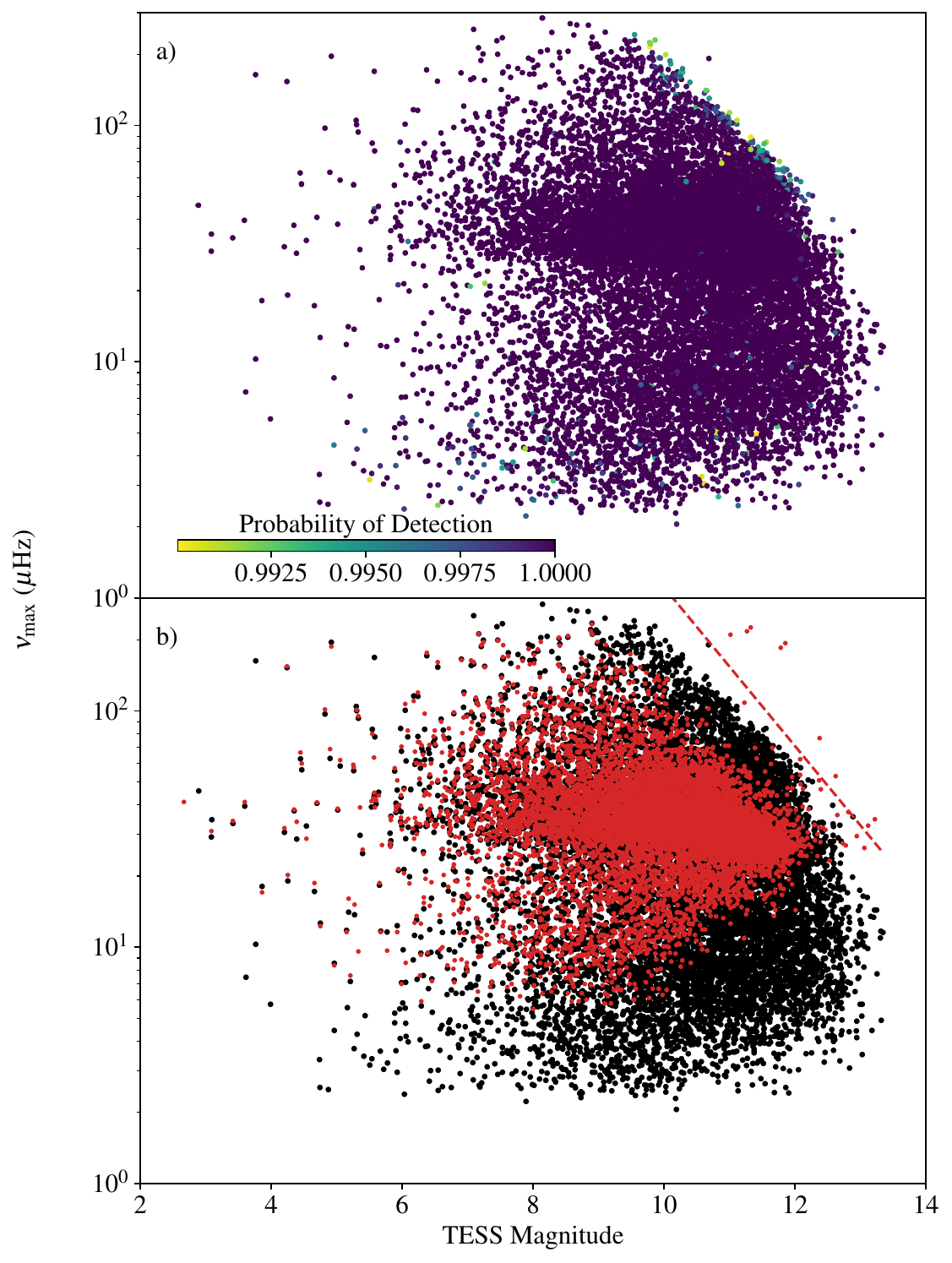}
    \caption{Frequency of Maximum power vs TESS Magnitude for all \fullosc{} stars in the sample. Top panel shows the stars colour coded by their detection probability. The bottom panel shows stars which were visually verified (black). The red points are the detections and the dashed line shows the empirical detection limit from \citep{honqlp}. }
    \label{fig:detection}
\end{figure}
\begin{figure*}
    \centering
    \includegraphics[width=0.8\linewidth]{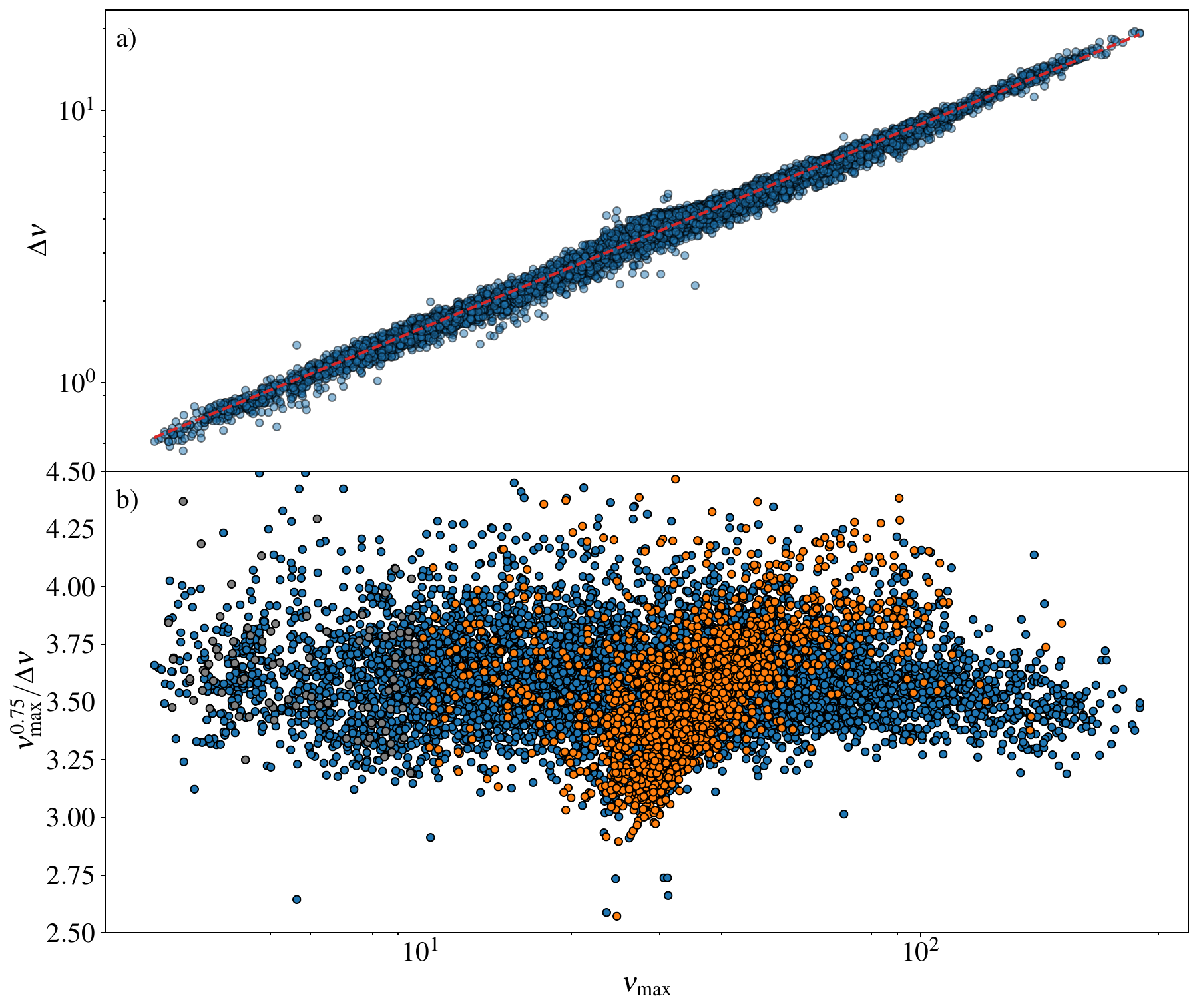}
    \caption{The correlation between \numax{} and \dnu{} for all \gooddnu{} stars with reliable \dnu{} measurements is presented in Panel a. The dashed line represents a power-law fit in linear space. Panel b provides the same data, with the 
y-axis serving as a proxy for mass. Blue points indicate \gooddnuRGB{} Red Giant Branch (RGB) stars, orange points denote \gooddnuRC{} Core Helium Burning (CHeB) stars, and grey points represent stars classified as ambiguous.}
    \label{fig:dnunumax}
\end{figure*}
In the frequency domain, solar-like oscillators show an excess of power whose shape is approximately Gaussian. Following \citet{Stello_2017, dennis22}, we visually examined the power density spectra of our \fullsample{} stars in logarithmic space and classified the detection of oscillations as either "yes",  "maybe" or "no" based on the characteristics of the power excess. This resulted in \visoc{} candidate oscillating stars (\yes{} yes and \maybe{} maybe). 

We next employed the \texttt{nuSYD} pipeline \citep{SREE24}, which is a simple and fast method to estimate \numax{} for this large sample of red giants. This method requires an approximate initial value for \numax{}, for which we used the correlation between Gaia absolute magnitude ($M_{G}$) and \numax{} for the 16,094 oscillating \kepler{} red giants reported by \cite{SREE24}. This correlation, shown in Fig. \ref{fig:mgvsnumax}, is analogous to a period-luminosity relation and allows the location of the power excess to be predicted approximately using only~$M_G$. We fitted a line to derive this relation:
\begin{equation}
\label{eqn1}
    \numax/\mu{\rm Hz} = 10^{0.52 M_G +  1.24}.
\end{equation} 
Although this relation is very approximate (scatter $\approx$ 40\,\%), it is sufficient for the \texttt{nuSYD} pipeline to converge to \numax{} if the star is oscillating. In using the pipeline, we measured the white noise at high frequencies, subtracted it from the power spectrum, and heavily smoothed the power spectra with a Gaussian kernel. We then reported the \numax{} corresponding to the maximum of the power excess ($P_{\rm peak}$) as the initial measurement of \numax{} for the \visoc{} candidate oscillators.
 
Having made initial measurements of \numax{}, it is important to verify whether these are genuinely solar-like oscillators. We followed the procedure outlined by \cite{chaplin11} and \cite{heyatl} to predict the empirical signal-to-noise ratio (SNR) and determine the probability of detection at the observed \numax{}. We used the \numax{}, the observational time span derived from our measurements, and effective temperature, $G$ magnitude, and distance from TIC to predict the total oscillation and granulation power inside the power excess envelope, and instrumental noise. Then, the signal-to-noise ratio (SNR) of the detected oscillations was calculated as the ratio between the total oscillation power inside the envelope and the total background power. Subsequently, we determined the probability ($p_{\rm det}$) that this SNR exceeds the predicted SNR. All these computations followed the equations provided by \cite{chaplin11} and \cite{campante2016}. 

Oscillation amplitude scales inversely with \numax{} \citep{huber_2011, mosser2012, yu18}. To represent the relationship with SNR, \citet{Stello_2017} introduced the \numax{} vs apparent brightness diagram. In such a diagram, the lowest SNR region appears at the top-right (with the highest SNR in the bottom-left), while the most distant stars are located in the bottom-right (with the least distant stars in the top-left). Figure~\ref{fig:detection}a shows \numax{} as a function of TESS magnitude, colour-coded by detection probability, for all \visoc{} stars in our sample. We find that approximately 96\,\% of these stars have a detection probability greater than 0.99. We therefore include only the \atlok{} stars out of these \visoc{} stars for our further analysis. 

From Fig.~\ref{fig:detection}b, we observe an increase in the number of low- and high-\numax{} stars at $\rm T_{mag}$ $>$ 8 (black circles) compared with \cite{honqlp} (red circles), which used a single sector of QLP lightcurves to detect oscillations (note that we detected oscillations in 97\% of the stars in that catalogue). This confirms that data from long-term TESS-based seismology can be used to probe less evolved and more distant stars. The dashed line in Fig.~\ref{fig:detection}b indicates the detection limit provided by \cite{honqlp}. Stars above this line, particularly those with high \numax{}, have amplitudes below the noise level, which makes detecting solar-like oscillations challenging. However, we visually inspected the pixel files and confirmed that no nearby bright stars existed within a radius of 4 pixels, so we retained these stars in our sample. By comparing with panel a, we see that stars near this line have slightly lower detection probabilities, yet close to unity. This confirms that we applied conservative criteria when identifying oscillating stars during the visual inspection of their power spectra.


\subsection{Blend analysis}

Due to the large pixel size of TESS (21.6 arcsec/pixel), a pixel could capture light from a companion star instead of the target star. If the companion star is an oscillating red giant, these oscillations would imprint their signatures on the wrong light curve. Previous studies have shown that contamination affects red giant oscillation spectra for stars fainter than a TESS magnitude of about 12 \citep{honqlp,dennis22}.
To ensure our sample of \atlok{} stars is free from false detections, we conducted a blending analysis following the procedure outlined by \cite{dennis22}. For each target star, we selected all stars within 150 arcsec and applied two criteria for identifying potential blends. We first calculated the Shape-Based Distance (SBD; \citealt{sbd2025}) between the power density spectra of the target star and all other stars in the list. We identified stars with an SBD value below 1 as potential blends, indicating that their power density spectra could be similar to that of the target star. For stars in this list with \numax{} values falling within 20\% of the target \numax{}, we measured the mean power within an envelope of width $0.33(\numax/\mu{\rm Hz})^{0.88}$ centred on \numax{}. If the power difference fell within 0.5\% of the entire sample, we flagged the star as contaminated.
We identified \nblend{} stars as contaminated which constitutes around 5.1\% of the total sample. We inspected the pixel files of a few of these stars and confirmed that they fall within the same or nearby pixel. We recognize that some of these stars may represent true physical associations. Therefore, rather than removing them from the sample, we assigned them a blend flag of 0 in Table \ref{tab:results}. After these steps, our sample contains \fullosc{} oscillating red giants.

\subsection{Global Asteroseismic parameters}

\label{sec:pysyd}
Two of the global asteroseismic quantities that can be measured from the frequency spectra of solar-like oscillations are the frequency of maximum power (\numax{}) and the large frequency separation (\dnu{}). These satisfy the approximate scaling relations \citep{ulrich1986, brown1991, kjb1995}, given by: 
\begin{equation}
\label{dnusacling}
    \frac{\dnu{}}{\dnusun{}} \approx \sqrt{\frac{\rho}{\rho_{\odot}} }
\end{equation}
and 
\begin{equation}\label{numaxscaling}
    \frac{\numax{}}{\numaxsun{}} \approx \frac{\rm g/\rm g_{\odot}}{\sqrt{\Teff/\Teffsun}}.
\end{equation}
By solving these equations, we can determine stellar mass ($M$), radius ($R$) and surface gravity ($g$) \citep{stello_2008_wire, kallinger2010}. 
\begin{equation}
         \label{eq:scalingmass}
    \frac{\rm M}{\msun} \approx \left(\frac{\numax{}}{\numaxsun{}} \right)^3 \left( \frac{\dnu}{\dnusun} \right)^{-4}\left( \frac{\Teff}{\Teffsun} \right)^{3/2},
    \end{equation}
    \begin{equation}
         \label{eq:scalingradius}
    \frac{\rm R}{\rsun} \approx \left(\frac{\numax{}}{\numaxsun{}} \right) \left( \frac{\dnu}{\dnusun} \right)^{-2}\left( \frac{\Teff}{\Teffsun} \right)^{1/2}
    \end{equation}
    and
     \begin{equation}
         \label{eq:scalingg}
    \frac{\rm g}{\gsun} \approx \left(\frac{\numax{}}{\numaxsun{}} \right) \left( \frac{\Teff}{\Teffsun} \right)^{1/2}.
    \end{equation}
 Here, $\numaxsun{} = 3090\,\pm\,30 $ \muHz{}, $\dnusun{} = 135.1\,\pm\,0.1$ \muHz{} and $\Teffsun{} = 5777\,K$ \citep{huber_2011, yu18}. To improve the \numax{} measurements from \texttt{nuSYD}, and to measure \dnu{} and associated uncertainties, we used the $\texttt{pySYD}$ pipeline \citep{huber2009, chantos22}. This involves modelling the stellar background at low frequency using Harvey functions and including a constant term for white noise. For the purpose of measuring \numax{}, we smoothed the background-corrected power density spectra with a Gaussian kernel, with a width of $\dnu{} * \max\left(2, 4\,(\numax/\nu_{\rm max,\odot})^{0.2}\right)$. We set a lower limit of 2\,\dnu{} instead of the $\texttt{pySYD}$ default value of 1\,\dnu{} to ensure that the envelope remains smooth at low \numax{} values \citep{SREE24}. We determined \numax{} as the peak of this smoothed oscillation envelope, while \dnu{} was derived from the autocorrelation of the background-corrected spectra. Note that $\texttt{pySYD}$ could not measure \dnu{}  for \pysydfail{} stars; we provide only the \numax{} value for these stars. 

To verify the quality of the \dnu{} measurements, we used the neural network developed by \cite{claudiya22}. The network uses \numax{}, \dnu{} and the background-corrected power density spectra from $\texttt{pySYD}$ to estimate the probability that the measured \dnu{} reproduces vertical ridges in the collapsed \echelle{} diagram. We first removed false-positive measurements that could reproduce vertical ridges, using the \numax{}--\dnu{} relation \citep{dennis2009}. Stars with a probability above 0.5 were considered reliable and flagged as good \dnu{} in the catalogue (see Table~\ref{tab:results}). This resulted in a sample of \gooddnu{} stars, or 56\% of the sample, with properly measured \dnu{}. Figure \ref{fig:dnunumax}a shows the \numax{}--\dnu{} diagram for these \gooddnu{} stars. Fitting a power law of the form $\alpha\left(\numax{}/\mu\textrm{Hz}\right)^{\beta}$ gives $\alpha = 0.282\pm0.003$ and $\beta = 0.750\pm0.003$. As expected, they follow the well-known relation between \numax{} and \dnu{} first shown by \citet{hekker2009} and \citet{dennis2009}.

  



\begin{figure}
    \centering
    \includegraphics[width=\linewidth]{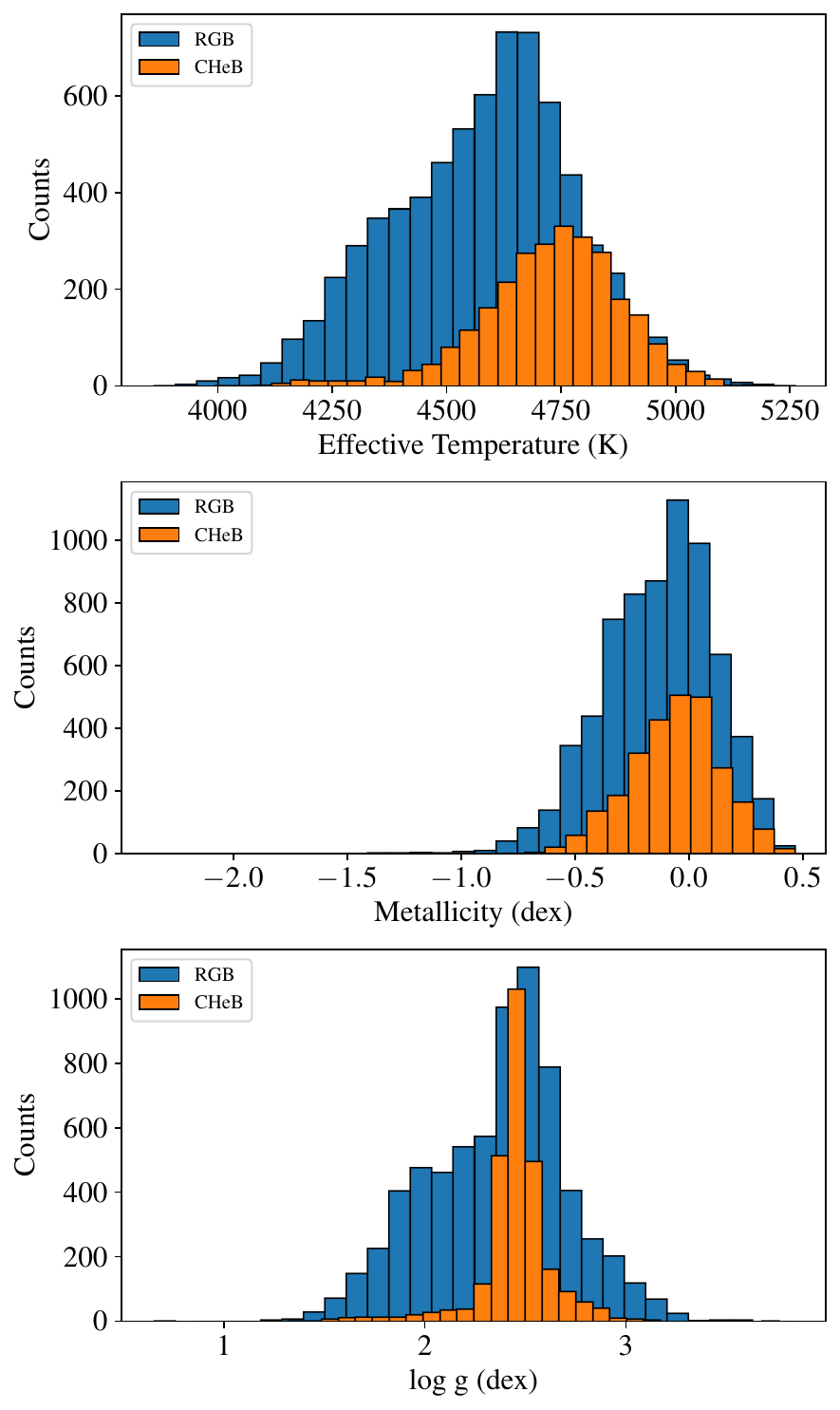}
    \caption{Distribution of stellar spectroscopic parameters from \citep{yu23} for our sample, red for \gooddnuRGB{} RGB and blue for \gooddnuRC{} CHeB stars. Panel a shows the effective temperature, panel b shows the metallicity and panel c shows \logg. }
    \label{fig:stellarparams}
\end{figure}
\begin{figure*}
    \centering
    \includegraphics[width=0.75\linewidth]{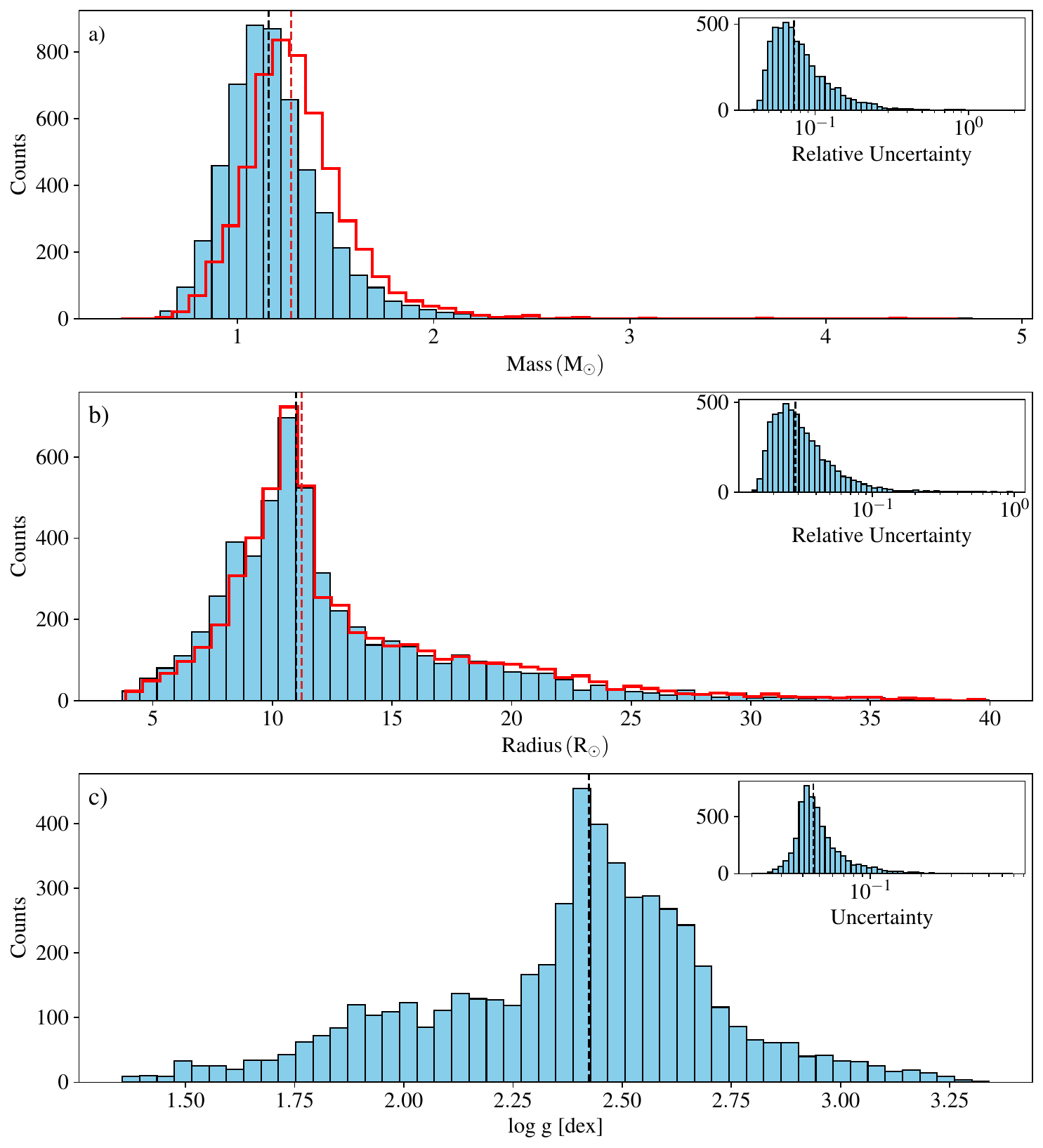}
    \caption{Distribution of Mass (top panel), Radius (middle panel) and log g (bottom panel) for the \gooddnu{} stars which has reliable measurement of \dnu{}. The insets show the uncertainty in stellar parameters. The red histograms show the distribution of uncorrected mass and radii. The dashed vertical lines show the median values of the distributions. }
    \label{fig:massradiuslogg}
\end{figure*}
\renewcommand{\arraystretch}{1.3}
\begin{table*}
\centering
\addtolength{\tabcolsep}{20pt}
    \caption{Columns in our catalogue of asteroseismic parameters for \visoc{} oscillating red giants in the TESS CVZs. Available at CDS via anonymous ftp to cdsarc.u-strasbg.fr (130.79.128.5) or \url{https://cdsarc.cds.unistra.fr/viz-
bin/cat/J/MNRAS}. Users should review the associated flags when utilizing the catalogue. }
    \begin{tabular}{lcc}
        \hline
        \hline
        Label & Description  \\
        \hline
        \hline
        TIC ID & TESS Input Catalog ID  \\
        GAIA ID & Source ID from GAIA DR2 \\
        ELON & Ecliptic Longitude (deg)  \\ 
        ELAT & Ecliptic Latitude (deg) \\
        Tmag & TESS magnitude \\
        $\rm N_{sectors}$ & Number of TESS sectors  \\
        Time span & Effective time span of TESS light curve (days) \\
        $\numax{}$ & Frequency of Maximum Power (\muHz{}) \\
        $e_{\numax{}}$ & Uncertainty in Frequency of maximum power (\muHz{}) \\
        $\dnu{}$ & Large Frequency Separation (\muHz{}) \\
        $e_{\dnu{}}$ & Uncertainty in Large Frequency Separation (\muHz{}) \\
        good  $\dnu{}$ flag & Flag indicating the reliable \dnu{} (1 =  reliable and 0 = unreliable)\\
        $P_{\dnu{} }$ & \dnu{} Vetting probability  \\ 
        Ev state & Evolutionary state ( 1 = RGB ,  2 = CHeB and 0 = ambiguous) \\
        Ev state score & Evolutionary state prediction score \\ 
        $p_{\rm det}$ & Detection probability \\ 
        Blend flag & Flag indicating whether the star is a potential blend ( 0 = a potential blend and 1 = non blend)  \\
        $\rm T_{\rm eff}$ & Effective Temperature (K) \\
        $\mathit{e}_{T_{\rm eff}}$ & Uncertainty in Effective Temperature (K) \\
        $\rm [M/H]$ & Metallicity (dex) \\
        $\mathit {e}_{\rm [M/H]}$ &Uncertainty in Metallicity (dex) \\
        \logg{} (spec)  & Spectroscopic Surface gravity (dex) \\
        $e_{\logg{}}$ (spec) & Uncertainty in Spectroscopic Surface Gravity (dex) \\
        spec source & Source of Spectroscopic data (1 = APOGEE/GALAH/RAVE and 2 = XP Spectra)\\\
        $\rm Mass_{\rm uncorrected}$   & Uncorrected Mass (\msun{}) \\
        $\rm Radius_{\rm uncorrected}$   & Uncorrected Radius (\rsun{}) \\
        \fdnu{} & Correction factor for \dnu{} scaling relation \\
        Mass & Stellar mass (\msun{}) \\
        $e_{\rm Mass}$ & Uncertainty in stellar mass (\msun{}) \\
        Radius & Stellar Radius (\rsun{}) \\ 
        $e_{\rm Radius}$ & Uncertainty in stellar radius (\rsun{}) \\
        \logg{} (seismic) & Surface Gravity (dex) \\ 
        $e_{\logg{}}$ (seismic) & Uncertainty on surface gravity (dex) \\
        Gold sample  & Whether member of gold sample (1 = member and 0 = non member) \\

        \hline
        \hline
    \end{tabular}
    \label{tab:results}
\end{table*}
\subsection{Determination of evolutionary states}
Based on their core properties, red giants can be classified into red giant branch (RGB), which burn hydrogen in a shell surrounding the core, and core-helium-burning stars (CHeB) (see reviews by \citealt{hekker2017, noels2025}). Oscillation frequencies reflect the properties of the core and envelope, and the period spacing of dipole ($l=1$) modes can be used to distinguish between RGB and CHeB stars \citep{bedding2011, mosser2014}.

We used the Convolutional Neural Network (CNN) by \citet{hon2018} to classify red giants based on the characteristics of RGB and CHeB stars in folded power spectra. The CNN uses images of the folded background-corrected power spectra and \dnu{} values to predict a score, using a network trained on labels from \kepler{} red giant data \citep{elsworth2019}. For each input, it predicts a score between 0 and 1 for the star to be a CHeB star. In addition to this score, the network also outputs the aleatoric uncertainty, related to the error in prediction due to noise in the data \citep{gal2014}. We classified stars with a score below 0.5 as RGB and those with a score above 0.5 as CHeB.
Figure \ref{fig:dnunumax}b shows the results from the neural network classifier for stars with the good \dnu{} flag = 1 (see Sec. \ref{sec:pysyd}). Among this, \gooddnuRGB{} stars were classified as RGB and \gooddnuRC{} stars as CHeB based on this score. We note that a small fraction of stars are classified as CHeB at lower \numax{} values. Given that low-\numax{} stars require longer observations and that core-helium burning is not expected at those \numax{} values, we classify them as “ambiguous” \citep{dennis2013,hon2018}.


\subsection{Stellar spectroscopic parameters}

Accurate determination of stellar mass, radius, and surface gravity requires reliable spectroscopic estimates of effective temperature and preferably also metallicity. 
Several spectroscopic surveys provide parameters for most stars in our sample. The Apache Point Observatory Galactic Evolution Experiment (APOGEE) produces stellar parameters, including surface abundances, for thousands of red giants mainly in the northern hemisphere, to study chemical composition and abundances \citep{abdrfapogee17}. Galactic Archaeology with Hermes (GALAH) targets nearby stars in the southern hemisphere to study the architecture and chemistry of the Galactic neighbourhood \citep{galah2015}. The RAdial VElocity experiment (RAVE), whose primary aim was to collect radial velocities for southern dwarfs and early giants, also provided stellar parameters for abundance studies \citep{rave2020}. Because surveys derive atmospheric parameters (effective temperature, metallicity, and surface gravity) using different methods, using parameters from different surveys can introduce systematic effects on different parameters and therefore it is essential to bring them to a uniform scale of measurement using proper calibration.


Here, we use stellar parameters from \citet{yu23} to calculate stellar mass, radius, and surface gravity. This catalogue provides effective temperatures and metallicities derived from APOGEE, RAVE, and GALAH spectra, which were homogeneously calibrated using common stars across the three catalogues and Gaia EDR3 parallax data. For the \xgbgood{}  stars with good \dnu{} that are not in the \citet{yu23} catalogue, we used the stellar parameters from \cite{Andrae_2023}. These are the stellar parameters derived from the Gaia XP spectra \citep{carrasco2021} using the XGBoost algorithm \citep{chen2016} which was trained using APOGEE spectroscopic parameters. This ensures that stars with reliable \dnu{} have stellar parameters in our catalogue. We compared the spectroscopic parameters with the corresponding parameters from \cite{yu23}, and found that the median offset between them was less than 0.1\%. As the spectroscopic parameters from \cite{Andrae_2023}  do not have uncertainties, we adopt an uncertainty of 2.8\% on effective temperature, 0.1 dex on metallicity and 3.8\% on \logg{}. Figure \ref{fig:stellarparams} presents the stellar parameters of \gooddnu{} stars in our sample. The distributions align with the general parameter ranges observed for red giants \citep{apokasc2,apokasc3}.


\begin{figure*}
    \centering
    \includegraphics[width=0.75\linewidth]{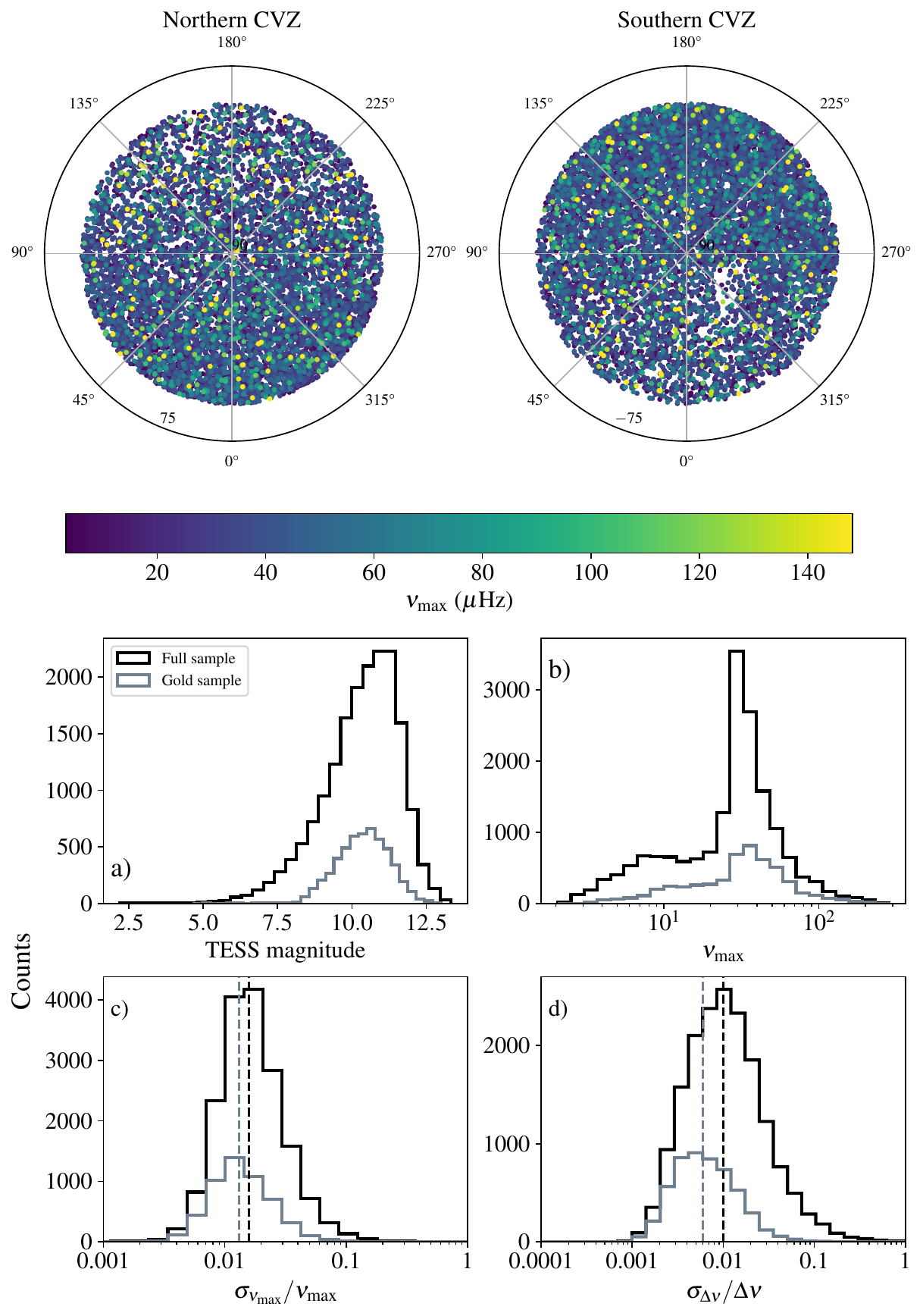}
    \caption{Properties of the stars in our final sample. The top two diagrams show the distribution of \fullosc{} oscillating red giants in northern CVZ and southern CVZ. Panel a shows the distribution of TESS magnitude, panel b shows the distribution of \numax{}, panel c shows the fractional uncertainty on \numax{} and panel d shows the fractional uncertainty on \dnu{}. The histograms in black corresponds to the full sample of red giants and grey corresponds to gold sample. }
    \label{fig:distsample}
\end{figure*}
\subsection{Determination of \fdnu{}}

Previous studies have identified a systematic offset between stellar parameters derived using scaling relations and those determined by full asteroseismic modelling \citep{White_2011, mosser2013b, guggen2017, asfgrid16, yaguangssurface23}. To address this offset, two correction parameters, \fdnu{} and \fnumax{}, were included in the scaling relations \citep{asfgrid16}. The corrected scaling relations are expressed as follows:

\begin{equation}
         \label{eq:correctedscaling}
    \frac{\rm M}{\msun} = \left(\frac{\numax{}}{\fnumax{}\numaxsun{}} \right)^3 \left( \frac{\dnu}{\fdnu{}\dnusun} \right)^{-4}\left( \frac{\Teff}{\Teffsun} \right)^{3/2}.
    \end{equation}
    \begin{equation}
         \label{eq:scalingradius2}
    \frac{\rm R}{\rsun} = \left(\frac{\numax{}}{\fnumax{}\numaxsun{}} \right) \left( \frac{\dnu}{\fdnu\dnusun} \right)^{-2}\left( \frac{\Teff}{\Teffsun} \right)^{1/2}.
    \end{equation}
     \begin{equation}
         \label{eq:scalingg2}
    \frac{\rm g}{\gsun} = \left(\frac{\numax{}}{\fnumax{}\numaxsun{}} \right) \left( \frac{\Teff}{\Teffsun} \right)^{1/2}.
    \end{equation}
\begin{figure*}
    \includegraphics[width=\linewidth]{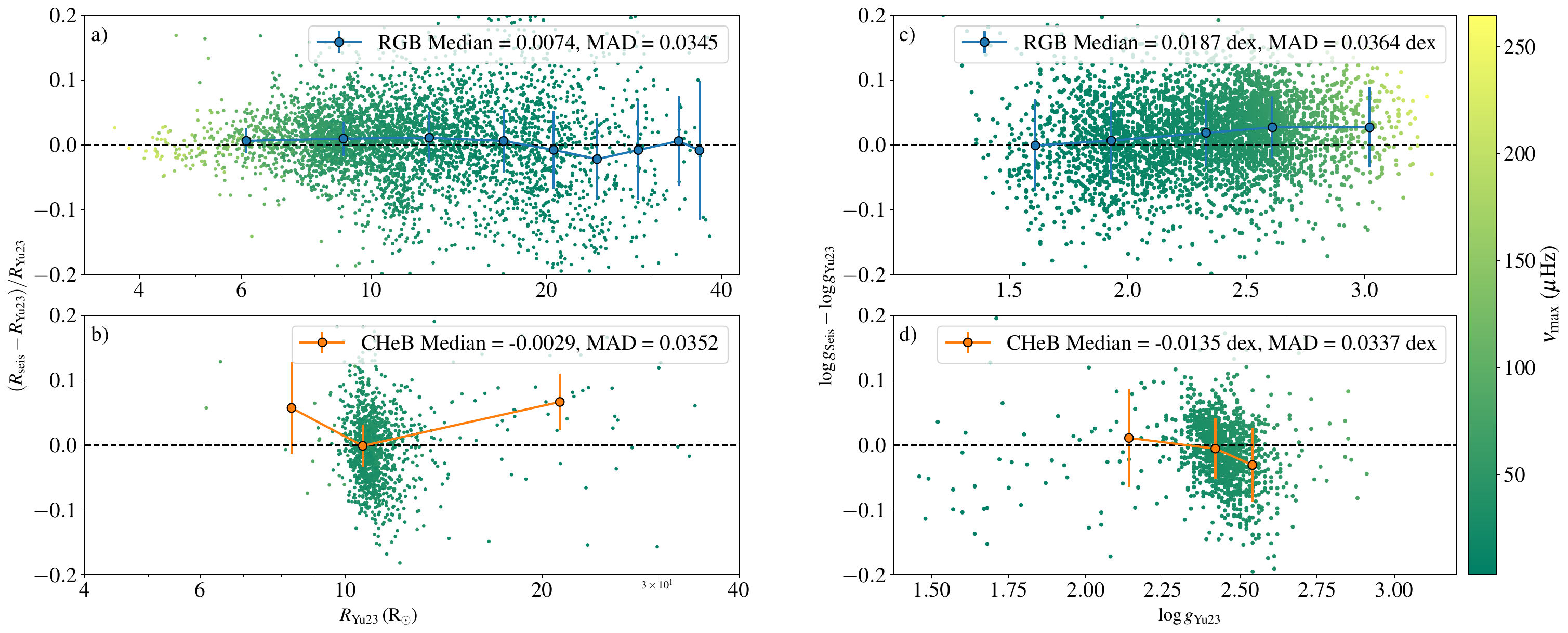}
    \caption{Comparison of asteroseismic radii with spectroscopic parameters, for the stars in gold sample. Panel a shows the comparison of GAIA radii for RGB stars and panel b shows the comparison for CHeB stars, both colour-coded by \numax{}. The blue and orange lines show the median offset between asteroseismic radii determined using \texttt{asfgrid} \citep{asfgrid16} corrections with Gaia radii for RGB and CHeB stars, respectively. The error bars show the median absolute deviation (MAD) per bin. The right-hand panels compare surface gravities derived from asteroseismology and those inferred from \citet{yu23}, for RGB stars (panel c) and CHeB stars (panel d), respectively. }
    \label{fig:gaia}
\end{figure*}
The parameter \fdnu{} accounts for the inexactness of eq. \ref{dnusacling}, which can be calculated from theoretical stellar models. For RGB and CHeB stars, we calculated the corrections using \texttt{asfgrid} \citep{asfgrid16, asfgrid22}. This procedure used the uncorrected mass, effective temperature, metallicity, and evolutionary states as inputs, interpolated the nearest stellar model grids, and derived \fdnu{}. Using the effective temperatures and metallicities from spectroscopy, we estimated the \fdnu{} parameter for \gooddnu{} stars that have reliable \dnu{} measurements. We verified that these values show clear trends with effective temperature and metallicity, as shown by \cite{asfgrid16}. Following \citet{apokasc3, aamanda2025}, we assumed an uncertainty of 0.005 for the \fdnu{} values \footnote{Note that some authors \citep[e.g.,][]{apokasc3, liagre2025} adopt a definition for \fdnu{} that is the inverse of that used here and by \texttt{asfgrid}.}.

Note that this approach does not include the standard surface term corrections to model frequencies, which also affects \dnu{} from models \citep{kj2008, ballandgizon2014, yaguangssurface23}. \cite{yaguangssurface23} parametrized the surface effect-corrected \fdnu{} as a function of \Teff{} and [M/H] for RGB stars and has shown that asteroseismic radii derived using these corrections agree well with Gaia radii. However, for CHeB stars,  \cite{schimak26} demonstrated that including surface corrections to \fdnu{} does not improve the asteroseismic masses and requires further studies to quantify it. To maintain uniform parameter values for the analysis, we chose not to calculate the masses and radii using these surface corrections. 

The value of \fnumax{} quantifies the inexactness of eq. \ref{numaxscaling}. Because \fnumax{} lacks a robust theoretical prediction from stellar models, its value is determined empirically from observations and theoretical models. Recent studies \citep{yaguang2022, huber2024,  jens2024, mia2025} identified a metallicity dependence in this correction factor. Currently, corrections are derived empirically by comparing asteroseismic radii with Gaia radii. \cite{apokasc3} fitted a polynomial to the ratio of \fdnu{}-corrected asteroseismic radii to Gaia radii and used this relation to account for discrepancies between asteroseismic and Gaia-derived radii. \cite{Marasco_2025} adopted a single value of \fnumax{} for all stars, defined as the median ratio between asteroseismic and Gaia radii. While early studies assumed a fixed value of unity for this correction term, more recent works calibrated it using Gaia measurements. As a reliable framework for \fnumax{} is not yet established, we assume \fnumax{} = 1 to compute stellar masses and radii. 

%




\subsection{Calculation of stellar parameters}
By combining our measurements of \numax{} and \dnu{} with the effective temperatures, and including the correction \fdnu{}, we calculated the stellar masses and radii using the scaling relations. We determined the uncertainties on stellar parameters by propagating the measurement uncertainties of \numax{}, \dnu{}, and effective temperatures. 
Our results, alongside the systematic uncertainties, are provided in our catalogue (see Table~\ref{tab:results}). Additionally, we provide uncorrected masses and radii (setting $\fdnu=1$) to facilitate studies aimed at calibrating the scaling relation, as these values represent the raw, uncalibrated outputs.

We present the distribution of stellar parameters in Figure \ref{fig:massradiuslogg}, which includes data for all stars with a good \dnu{} flag set to 1 and homogeneous stellar parameters. The distribution spans a range of stellar masses, with a median mass of 1.17\,\msun{} (corrected) and 1.23\,\msun{} (uncorrected). The uncertainties in mass measurements have  a median relative value of 7.5\%. The measured radii span the expected range, with a median relative uncertainty of 2.8\%, and align with the trends observed for red giants in earlier catalogues \citep{yu18, hon22}. The surface gravities have a median value of 2.45 dex, with a median uncertainty of 0.01 dex, showing the relatively high precision offered by asteroseismic surface gravities. Note that the quoted uncertainties on the stellar parameters do not include systematic uncertainties arising from surface effects and from \fnumax{} calibration. 

\section{Analysis} \label{sec:highlight}

\subsection{Properties of the catalogue}
We present a catalogue of \fullosc{} oscillating red giants in the TESS CVZ, with consistent and homogeneous measurements. Figure \ref{fig:distsample} (upper two plots) shows the sky distribution of the stars in our sample. In total, the NCVZ contains \fullsamplenorth{} oscillating red giants, while the Southern CVZ contains \fullsamplesouth{}. The distribution of the detections on the sky does not show any significant features, apart from a gradient resulting from the higher number of stars towards the Galactic plane. 

To ensure reliable measurements, we restrict our analysis in the following sections to \goldall{} stars (hereafter referred to as the gold sample) with probability of detection $\geq$ 0.99, blend flag set to 1, good \dnu{} flag set to 1, evolutionary–state scores $\leq$ 0.1 (clear RGB) or $\geq$ 0.9 (clear CHeB), and stars where masses and radii were calculated using spectroscopic parameters from \cite{yu23}. 
The full sample spans $\rm T_{mag}$ from 2 to 13, while the brightest star in the Gold Sample has a $\rm T_{mag}$ of 4.8 (Fig. \ref{fig:distsample}a). In Fig. \ref{fig:distsample}b, we see a slight excess of stars around \numax{} of 9\,\muHz{}, possibly due to the Asymptotic Giant Branch stars (AGB; \citealt{guillame2022}). Figures~\ref{fig:distsample}c and d show the precision of \numax{} and \dnu{} for our stars. The median uncertainty in \numax{} is \fullnumaxprec{}\% for the full sample and \goldnumaxprec{}\% for the gold sample, achieving a precision comparable to \kepler{} data \citep{yu18}. For \dnu{}, we achieved a median precision of \fulldnuprec{}\% for the full sample, improving to 0.6\% for bright stars ($\rm T_{mag}$ $< 8$) and 1.3\% for stars with $\rm T_{mag}$ $\sim12$. In the gold sample, the median \dnu{} precision is \golddnuprec{}\%, demonstrating its higher quality relative to the full sample. This translates to a precision of 7.2\,\% in mass, 2.6\,\% in radius and 0.01 dex in \logg{}, comparable to the precision from 4-yr \kepler{} data.

\subsection{Comparison with Gaia radii and spectroscopic surface gravities}


We compared our asteroseismic radii with those from Yu et al. (2023), which were derived from bolometric fluxes, distances, and effective temperatures. Yu et al. (2023) obtained the bolometric fluxes by integrating best-fitting model atmospheric spectra \citep{kurucz1979, bosz2017}. These spectra were fitted to spectral energy distributions (SEDs) constructed from Gaia EDR3 and 2MASS photometry \citep{rielo2021, 2mass}. The distances were adopted from Bailer-Jones et al. (2021), which account for the Gaia EDR3 parallax zero-point offset \citep{lindegren2021}. Effective temperatures were taken from the APOGEE \citep{abdrfapogee17}, GALAH \citep{galah2015}, or RAVE \citep{rave2020} surveys. 

Figure \ref{fig:gaia} shows the comparison, where the fractional difference between asteroseismic and Gaia radii is plotted for both RGB and CHeB stars. The radii calculated in this work exhibit almost a constant median offset of 0.7\% with a scatter of 3.4\% up to $\approx$\,15\,\rsun{} for RGB stars. We observe that the scatter increases further for stars at more advanced evolutionary stages. For CHeB stars, we see an offset of 0.3\% with a scatter of 3.4\%. These observed scatter between asteroseismic and Gaia radii are smaller than the typical fractional uncertainty of 7\% in Gaia radii \citep{yu23}, suggesting that the observed scatter is likely dominated by measurement errors in Gaia radii. This result is smaller than the previous reports of an offset of approximately 4\% between asteroseismic and Gaia radii \citep{Berger_2018, Zinn2021, Zinn2023, yu23}. 

We also compared the surface gravities derived in this work with those obtained by \cite{yu23}. For RGB stars, we see an overall offset of $0.02\pm0.04$; however, the offset gradually increases as the stars become less evolved (Fig.~\ref{fig:gaia}c). For CHeB stars (Fig.~\ref{fig:gaia}d), the offset is smaller ($-0.01\pm0.03$), but there is a slight trend in the offset as a function of \logg{}. These offsets between asteroseismic and spectroscopic surface gravities agree well with the previous studies \citep{thygsen2012, morelmigilo2012, yu23, apokasc3}.

\subsection{Comparison with previous TESS studies}
\begin{figure}
    \centering
    \includegraphics[width=\linewidth]{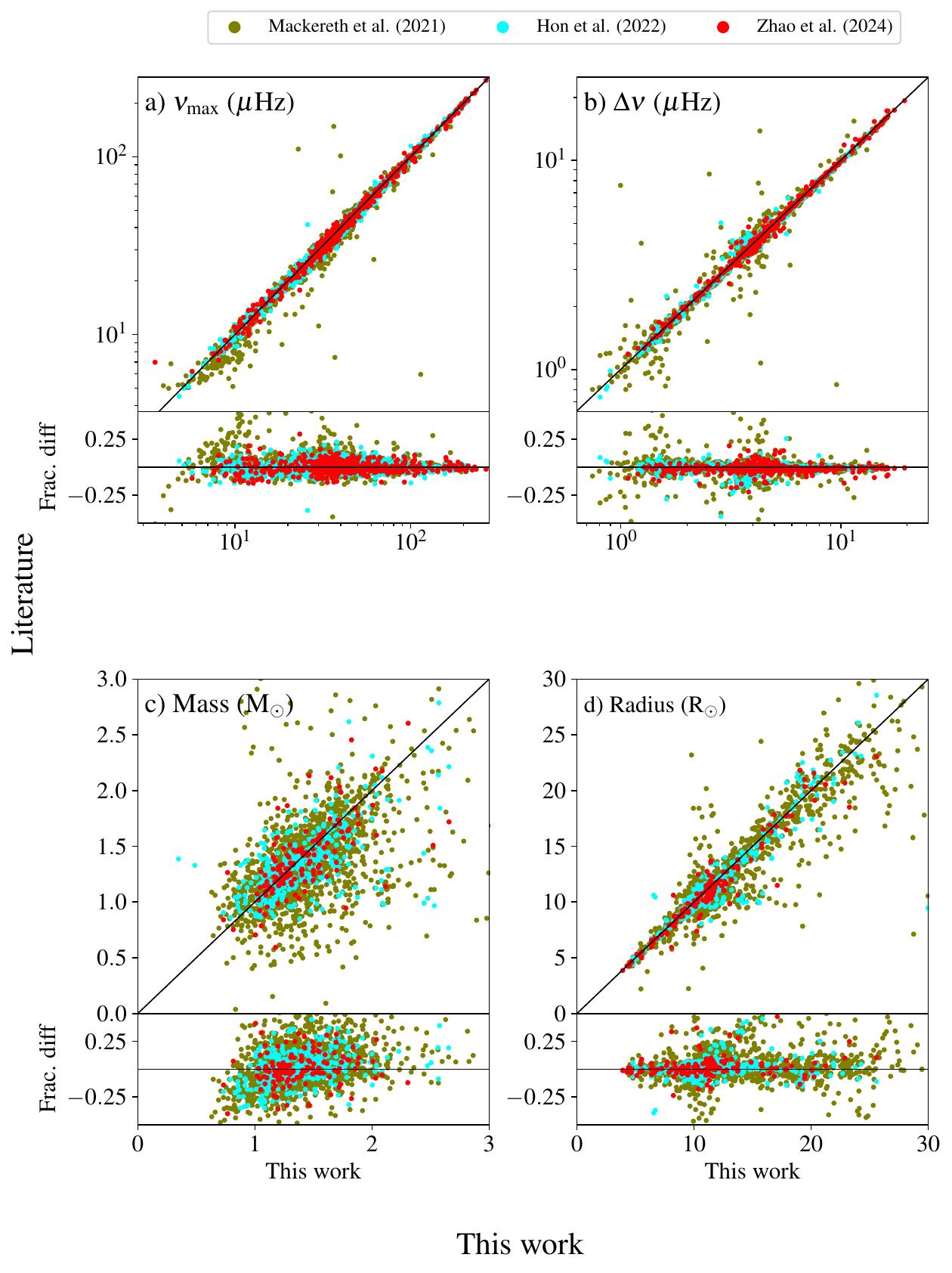}
    \caption{Comparison of asteroseismic and stellar parameters  with measurements from previous studies. Red points corresponds to values from \citet{z24}, olive points from \citet{m21} and cyan points from \citet{hon22}. Panel a shows the comparison of \numax{} values and panel b shows the comparison of \dnu{} values. panel c and d represents mass and radius, respectively. The fractional difference is also shown on the bottom axis. }
    \label{fig:comparisoncatalogues}
\end{figure}
Three previous studies with TESS have carried out asteroseismology of red giants in the Continuous Viewing Zones. \citet{m21} conducted a study of red giants in the TESS Southern CVZ from the first year of the mission (Sectors 1--13), which included a region wider than the true CVZ. Using asteroseismic parameters from the Birmingham Asteroseismology Pipeline (BHM) combined with effective temperatures from APOGEE DR17, they derived the masses and radii of approximately 15,000 stars. They then refined the stellar parameters derived from asteroseismic scaling relations using Bayesian optimization with priors from stellar models. \cite{hon22} carried out an asteroseismic analysis of 2,600 bright stars in the CVZ that are members of the HD catalogue. Using data from TESS Sectors 1--39, they measured \numax{} and \dnu{} for 1,700 stars. Their mass and radius estimates were refined using \texttt{asfgrid} corrections. Finally, \cite{z24} used two-minute cadence light curves from TESS Sectors 1--60 to determine \numax{} and \dnu{} for subgiants and red giants. Using stellar atmospheric parameters from Gaia Radial Velocity Spectrometer (RVS) spectra, they produced a catalogue of masses and radii for 8,000 solar-like oscillators, some of which are in the CVZs.


We cross-matched our catalogue, which is based on Sectors 1--\lastsector, with these three catalogues of solar-like oscillators to evaluate the level of agreement. Since different studies use varying stellar parameters and correction methods, we used the \numax{} and \dnu{} values from these studies combined with the effective temperatures from our study to derive stellar masses and radii. Figure \ref{fig:comparisoncatalogues} shows the comparison of asteroseismic and stellar parameters. The \numax{} values show excellent agreement, with a discrepancy of less than 1\% when compared with \cite{z24} and \cite{hon22}, and a median fractional difference of approximately 1.4\% when compared with \cite{m21}. This discrepancy may arise from \cite{m21} using a different pipeline to measure \numax{}. However, the \dnu{} values from all three studies show consistent agreement, with a variation of less than 0.5\%. The derived masses exhibit offsets of 2.2\% and 1.5\% when compared with \cite{m21} and \cite{hon22}, respectively. These offsets are within the median uncertainty of the mass measurements.
\begin{figure}
    \includegraphics[width=\linewidth]{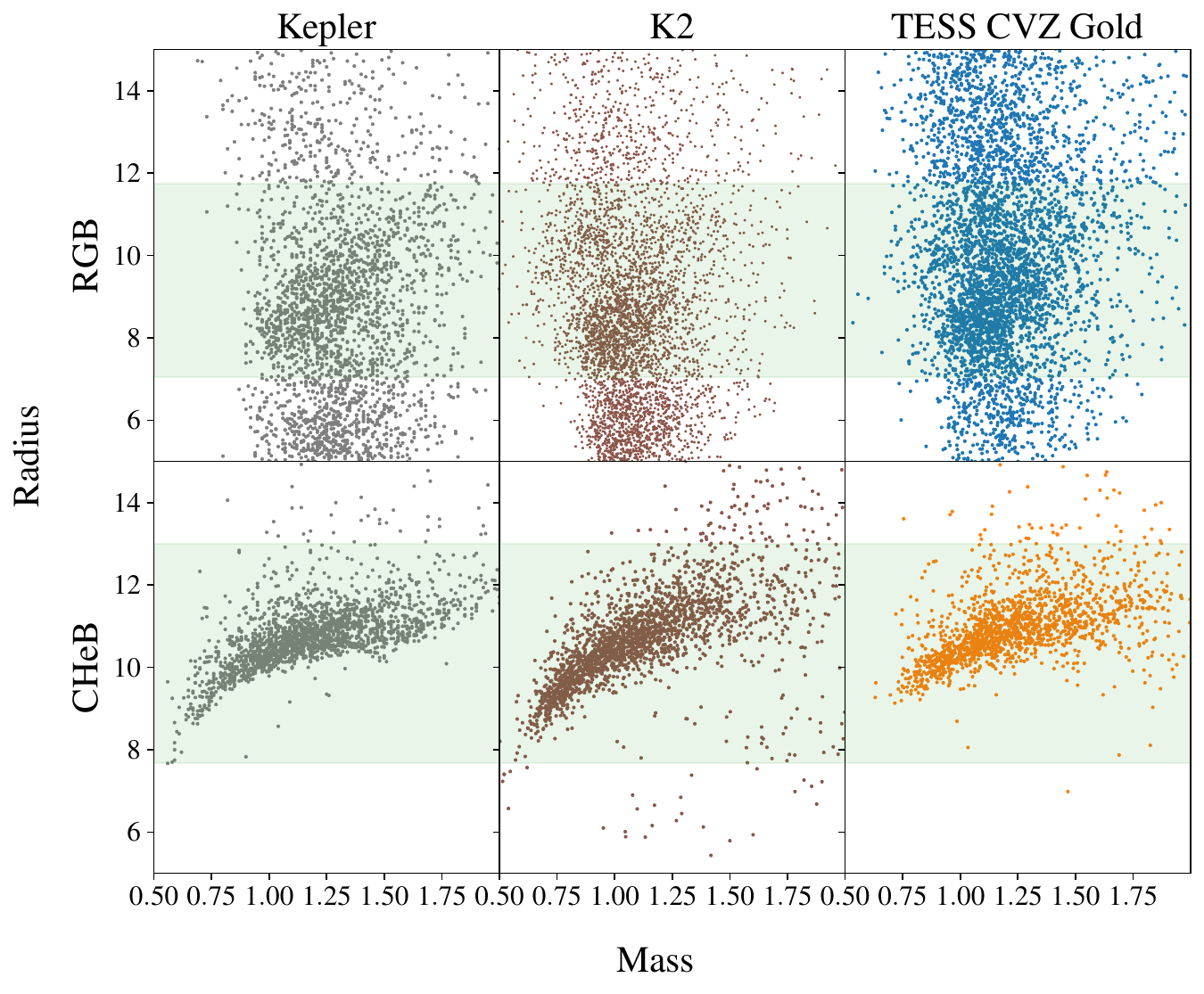}
    \caption{Comparison of the RGB bump and the red clump between \kepler{}, K2 and TESS. The top panel shows the mass--radius plots of the RGB stars and the bottom panel is for CHeB stars. The green shaded region is an arbitrarily chosen region to guide the eye for locating approximate locations of RGB bump and CHeB clump.}
    \label{fig:edges}
\end{figure}

\subsection{Comparison with \kepler{} and K2 red giants}
The asteroseismic properties of \kepler{} and K2 red giants have been studied extensively, significantly advancing our understanding of stellar evolution \citep{chalinmigilo2013, mosser2012, vrard2016, jackewiez2021, apok2, apokasc3, noels2025}. These properties serve as an important reference point for assessing stellar parameters derived from TESS CVZ observations. Although \kepler{} and K2 targets are located in different regions of the sky, the physical insights gained from their asteroseismic data have been invaluable for understanding red giants. During their ascent along the Red Giant Branch (RGB), stars undergo a brief phase where their luminosity decreases before increasing again. This period of slower luminosity evolution causes an accumulation of stars in a specific region of the observational parameter space, known as the RGB bump. Similarly, when low-mass stars ignite helium in their cores, they form a sharp transition in the Hertzsprung-Russell diagram, referred to as the Zero Age Helium-Burning (ZAHeB) edge \citep{jcd2015, hekker2017, girardi2016}. Precise asteroseismic mass--radius (M--R) diagrams from \kepler{} and K2 data have successfully reproduced these sharp features, providing key insights into different stages of stellar evolution and allowing a test of the accuracy of the asteroseismic scaling relations \citep{yaguang22}. In this section, we assess whether our TESS CVZ data reveal similar features and compare them with \kepler{} and K2 datasets \citep{apokasc3, apok2}.

The upper panel of Fig.~\ref{fig:edges} shows the mass--radius diagrams for RGB stars in the \kepler{}, K2, and TESS CVZ gold samples. In all three samples, the RGB bump occurs in the same region. However, we observe a deficit of low-mass RGB stars in the CVZ sample compared to \kepler{} and K2.
The bottom panel of Fig.~\ref{fig:edges} presents mass-radius diagrams for 1,110 core helium-burning (CHeB) stars in the gold sample. The zero-age CHeB sequence for the CVZ stars is well-aligned with the corresponding \kepler{} CHeB stars. However, comparing the precise location of the ZAHeB edge between \kepler{}, K2, and TESS requires more detailed analysis, which we will address in future work. Interestingly, we also identify a few very-low-mass core helium-burning stars in the CVZ sample that appear to have undergone post-mass-transfer events, consistent with discoveries in \kepler{} data by \citet{yaguangnature}.


\begin{figure*}
    \centering
    \includegraphics[width=0.9\linewidth]{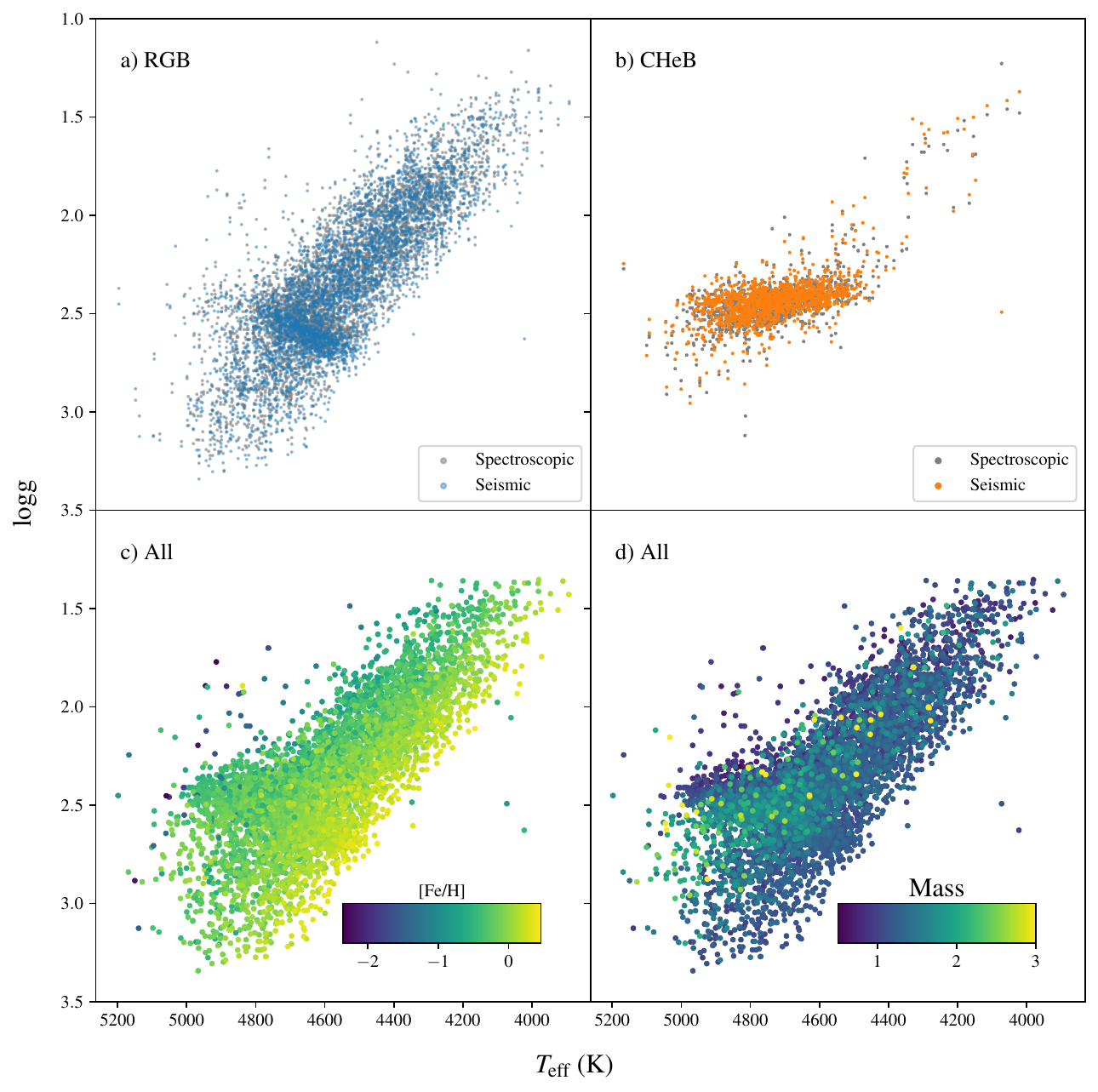}
    \caption{Kiel diagram of the \goldall{} stars in gold sample. The x-axis represents the effective temperature, while the y-axis corresponds to \logg{}. Panel a illustrates the asteroseismic \logg{} as blue points and the spectroscopic \logg{} as grey points for \goldRGB{} RGB stars. Panel b displays the Kiel diagram for \goldRC{} CHeB stars, represented in orange. Panels c and d show the Kiel diagram for all stars, colour-coded by metallicity and mass, respectively.}
    \label{fig:hr}
\end{figure*}

\subsection{Evolution in the Kiel diagram}

The asteroseismic \logg{} values derived in this work allow us to investigate fundamental stellar evolution and validate previous results. Previous studies have demonstrated that asteroseismic \logg{} provides smaller uncertainties compared to spectroscopic \logg{}, making it a more precise measurement \citep{morelmigilo2012, thygsen2012, hekker2013, apokasc1}. Such precise constraints help in determining accurate ages as well as calibrating spectroscopic data \citep{Gai_2011, Mészáros_2013, creevy2013, hekkergalac}. Figure \ref{fig:hr}a shows the Kiel diagram for RGB stars using our asteroseismic \logg{} and the spectroscopic \logg{} inferred by \citet{yu23}. We observe that the RGB bump is sharper with asteroseismic \logg{}, confirming the accuracy of asteroseismology. As previously observed with CHeB stars, the clump region is evident from asteroseismic \logg{}. However, we identify CHeB stars at \logg{} $<$ 2, a feature not observed in \kepler{} data. We used MIST tracks \citep{choi2016} to verify whether CHeB stars can exist at larger radii and found that their positions lie between the Helium flash and Core Helium Burning regions. The typical lifetime ratio of these stars is only 0.002 compared to the CHeB state, suggesting that $\sim3$ stars in our sample may be passing through the helium-flashing phase \citep{Bildsten_2012, sebastian2018}. Therefore, we conclude that CHeB stars at \logg{} $<$ 2 are presumably misclassified.


We can explore stellar evolution in the Kiel diagram, using \logg{} as a proxy for the luminosity of stars on the giant branch \citep{tayar2019, apokasc2, apokasc3}. According to stellar evolution, metallicity affects the locations of the Red Giant Branch (RGB) bump and the CHeB clump (also referred to as the Red Clump, RC) \citep{kwr2013, girardi2016, khan2018, apok2}. Figure \ref{fig:hr}c shows the Kiel diagram, with points colour-coded by metallicity. At a given \logg{}, we observe that each sequence shifts to higher temperatures as metallicity decreases, indicating that the positions of both the RGB bump and RC vary with stellar metallicity. 

In Fig.~\ref{fig:hr}d we colour-code the stars based on the asteroseismic mass derived in this work. In the RGB bump and in the CHeB clump, we observe that the location of the bump has a slight dependency on mass. Stars located below the clump have higher masses, confirming that they are secondary clump stars \citep{girardi2016}.


\begin{figure*}
    \includegraphics[width=0.9\linewidth]{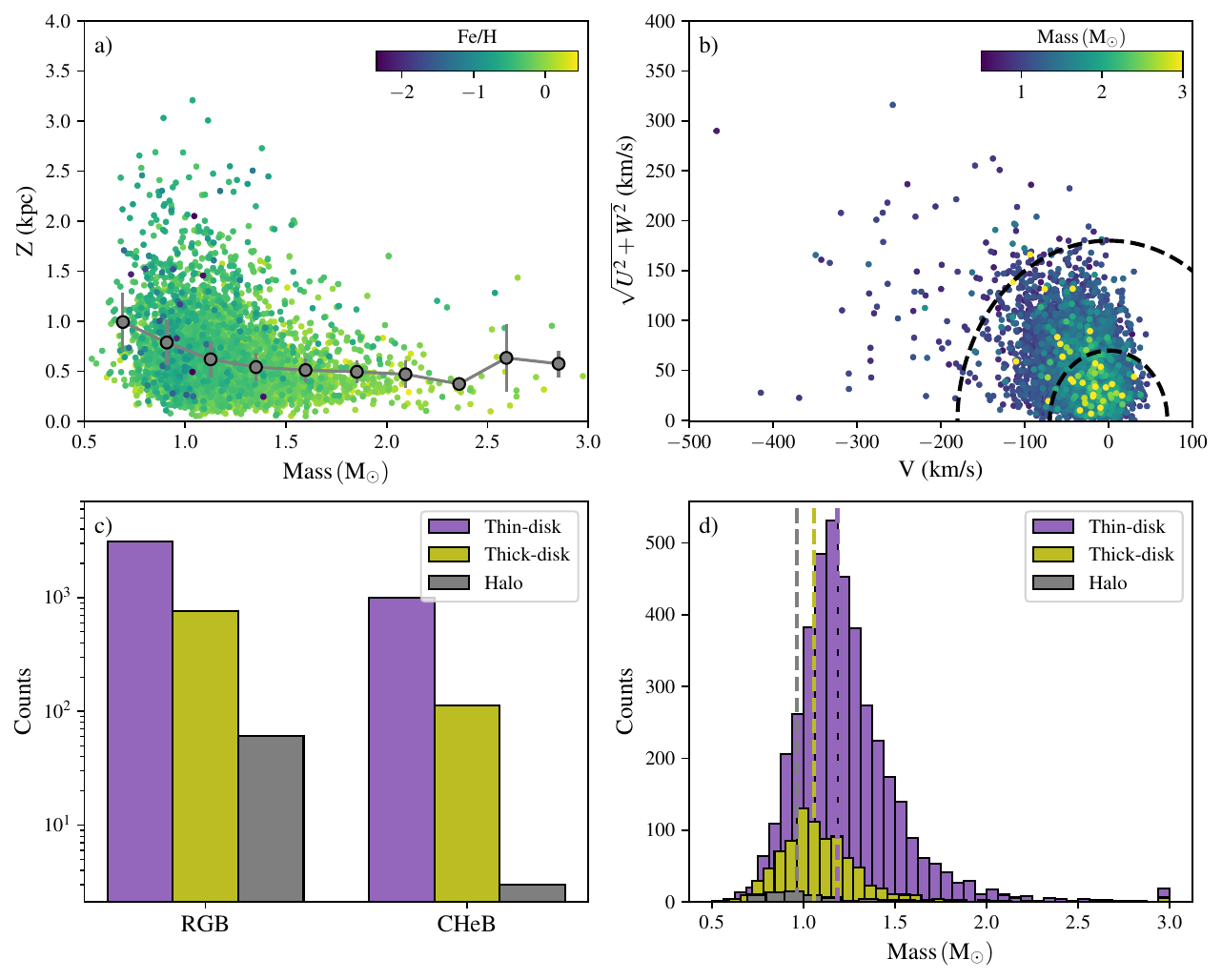}
    \caption{Analysis of the gold sample in Galactocentric coordinates. Panel a shows the positional distribution of all the oscillating stars, colour-coded by mass. The solid lines show the median $z$ per bin and error bars show the median absolute deviation (MAD). Panel b shows the Toomre diagram colour-coded by mass. The dashed semicircles show the boundaries from \citet{Mardini_2022} used to delineate different populations. Panel c shows the distribution of RGB and CHeB stars in different populations. Panel d shows the distribution of asteroseismic mass in these populations. The vertical dashed lines show the median mass in each of these populations.}
    \label{fig:galactic}
\end{figure*}

\subsection{Potential for Galactic archaeology}
To understand the distribution of stars in our sample within a Galactic context, we analysed their spatial properties in Galactocentric coordinates. Using distances from Gaia DR3 \citep{gaiadr3}, we calculate the static Galactocentric positions with the \texttt{astropy Galactocentric} module \citep{galactocentric}. In this system, the $x$-axis points from the Sun to the Galactic centre, the $y$-axis points in the direction of Galactic rotation, and the $z$-axis points toward the North Galactic Pole. We adopt a distance of 8.21 kpc from the Sun to the Galactic centre and a Cartesian velocity of 233.1 km/s \citep{mcmilan2017}. 

The vertical distance ($z$) from the Galactic plane correlates with stellar age; older stars tend to reside farther from the Galactic plane \citep{cassagrande2016, Ness_2016}. Figure~\ref{fig:galactic}a shows $z$ as a function of mass. We observe that the average stellar mass increases closer to the plane, confirming that younger stars are generally found closer to the Galactic plane \citep{chiappini1997, migilo2013, chiappini2015, cassagrande2016, migilo2017}. Additionally, metal-poor stars are known to reside farther from the Galactic plane, suggesting a metallicity dependence on vertical distance \citep{Carraro1998, soubiran20004, soubiran2008}. Figure \ref{fig:galactic}a demonstrates that metallicity decreases with increasing $z$, consistent with previous Galactic studies \citep{katz2011, bovy2012}.

 \begin{figure*}
    \centering
    \includegraphics[width=0.8\linewidth]{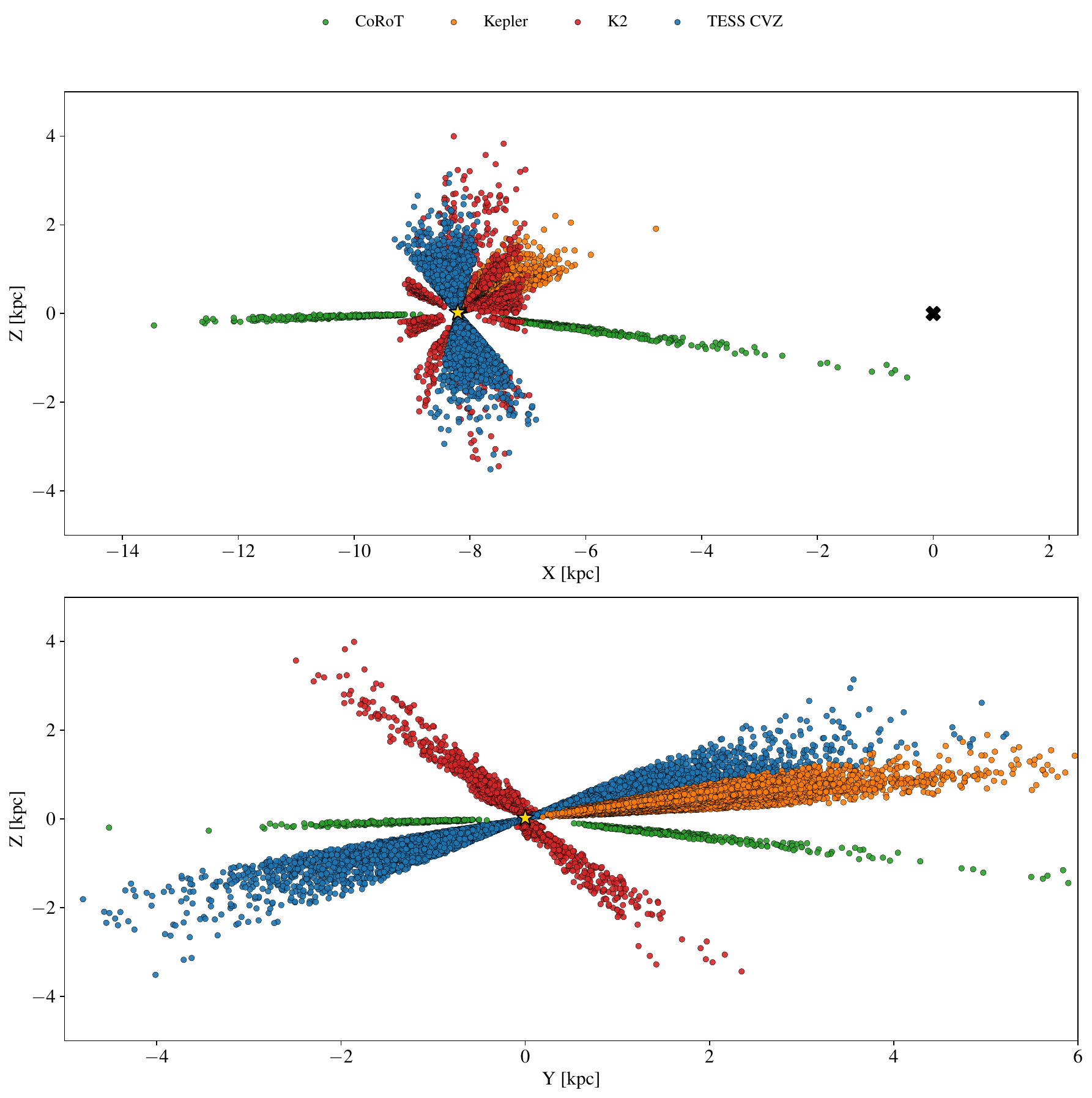}
    \caption{Projection of stars with asteroseismic ages from various space missions in Galactocentric coordinates. The top panel shows the vertical height from the plane as a function of distance from the Galactic centre, which is denoted using a cross symbol.  The bottom panel shows the distribution as viewed towards galactic centre. The orange points are \kepler{} data from \citet{apokasc3}, green points for \corot{} are from \citet{anders2017}, red points for K2 are from \citet{apok2} and blue points are for TESS CVZ, with good \dnu{} flag set to 1 from this work. The cross marks the Galactic center.}
    \label{fig:galproj}
\end{figure*}  

 Stars closer to the Galactic plane tend to exhibit higher velocities than those farther from the plane \citep{misha2013}. Using proper motions and radial velocities from Gaia DR3 \citep{gaiadr3}, we constructed the Toomre diagram with Galactocentric velocities ($U$, $V$, and $W$), shown in Figure \ref{fig:galactic}b. This diagram is widely used in kinematic studies to analyse the chemistry and dynamics of the Galaxy \citep{venn2004}. Figure \ref{fig:galactic}b shows the velocity distribution of red giants in our gold sample, where stars are colour-coded by asteroseismic mass. We classify the stars into three populations-thin disk, thick disk, and halo-using the total velocities from \citet{Mardini_2022}. We observe a gradient in stellar mass with the total velocity, arising as a consequence of mass-age-velocity relations \citep{migilo2013, misha2013}.  Based on this classification, our sample contains 4,115 stars in the thin disk, 870 stars in the thick disk, and 64 halo stars. Figure \ref{fig:galactic}c shows the evolutionary state of the red giants in each population. The total number of stars in each population decreases with distance from the Galactic plane \citep{read24}. However, we note a significant reduction in the number of CHeB stars within the halo population. Figure \ref{fig:galactic}d shows the mass distribution for each stellar population. The median mass decreases from the thin disk to the halo population, which has the lowest velocities. As stellar mass can be considered as a proxy for age \citep{migilo2013},  this decrease reflects the well studied age-velocity relation \citep{chiappini1997, misha2013, SilvaAguirre2018}.

Overall, this catalogue will be useful for calculating stellar ages \citep{camila2022, jamie2022}. A sample with a large number of red giants is valuable for studying the structure and formation of the Milky Way \citep{chiappini2015, hekkergalac, migilo2017}. Figure \ref{fig:galproj} shows the projection of stars with asteroseismic ages from different space missions. Previous missions, such as \corot{} and K2, focused on regions near the Galactic and ecliptic planes, respectively, while the TESS CVZ data emphasize the ecliptic poles. Combining the asteroseismic masses from this work with astrometric data from Gaia allows us to describe the spatial extent of the sample. When integrated with chemical abundances from spectroscopy, this information offers chemo-kinematic insights into the region \citep{apokasc3, k2gap, roberts24, Marasco_2025}.





\section{Conclusions}
In this work, we utilized the long-duration light curves available for the TESS Continuous Viewing Zones in Sectors 1--\lastsector{} to compile a catalogue of \visoc{} oscillating red giants. The catalogue provides homogeneously derived frequency of maximum power (\numax{}), large frequency separation (\dnu{}), evolutionary state classifications, asteroseismic masses, radii, and \logg{} values (see Table \ref{tab:results}).

 \begin{enumerate}
    \item We detected oscillations in \visoc{} red giants through manual inspection of TESS light curves. We verified that these stars are oscillating by comparing their SNR with expectations \citep{chaplin11, heyatl}. We found an increase in the number of stars with high and low \numax{} values at fainter magnitudes compared to previous studies that were based on fewer sectors. This finding suggests two key points: (1) the long-duration light curves from TESS enable the detection of bright red giants at greater distances, facilitating exploration of more distant regions of the sky; and (2) extended light curves reduce noise levels, allowing the detection of high \numax{} oscillations.    
    
    \item We measured \numax{} and \dnu{} for all the stars using \texttt{pySYD} \citep{huber2009, chantos22}, achieving a typical precision of 1.5\,\% and 1.0\,\%, respectively. We used the Neural Network framework by \citet{claudiya22} to assess the reliability of these measurements. For stars with a prediction score from Neural Network above 0.5, we found a typical median uncertainty of 0.6\,\% in \dnu{}, achieving a precision comparable to that of \kepler{} data.

    \item We classified the stars into Red Giant Branch (RGB) stars and core Helium-burning (CHeB) stars using the Convolutional Neural Network proposed by \citet{hon2018}. Stars with classification scores below 0.5 were labelled as RGB (denoted by 1), while those with scores above 0.5 were labelled as CHeB (denoted by 2). Contrary to previous studies, we observed CHeB stars with \numax{} values as low as 7\,\muHz{}. Although this result is unexpected, we do not rule out the presence of core Helium-burning stars at such low \numax{} values. We labelled these stars as ambiguous, and they warrant further detailed study in the future. 

    \item Using asteroseismic scaling relations, we derived precise masses, radii, and surface gravities for \gooddnu{} stars by combining asteroseismic data with spectroscopic parameters from APOGEE, GALAH, RAVE \citep{yu23}, and Gaia XP spectra \citep{Andrae_2023}. For more detailed analysis, we selected a subsample of \goldall{} stars, which we labelled the "gold sample". This sample has median precisions of 7.2\,\% for mass, 2.6\,\% for radius, and 0.01 dex for \logg{}.

    \item We find that the locations in the mass-radius diagram of the RGB bump and Zero Age Helium-Burning (ZAHeB) edge are consistent with in previous studies using \kepler{} and K2.  Additionally, we demonstrate that the \logg{} values derived in this work are more precise than those obtained from spectroscopic measurements, suggesting that they can serve as a reliable reference for calibrating existing spectroscopic pipelines.
    
    \item Combined with Gaia astrometric data, we find that stellar mass decreases with increasing Galactic height, consistent with the age-mass relation, where older stars tend to be less massive and dynamically heated to higher altitudes. Furthermore, the average metallicity decreases with increasing distance from the Galactic plane. Similar trends are observed in stellar velocities, indicating that the asteroseismic masses derived in this work can serve as a reliable proxy for stellar age.
\end{enumerate}


In summary, this study demonstrates that continuous long-duration photometry from TESS, combined with spectroscopy, enables precise mass estimation of large numbers of red giants. The large number of stars in this homogeneous sample could help to improve current ensemble asteroseismology methodologies and spectroscopic pipelines. Although a detailed analysis lies outside the scope of this paper, we highlight some key features relevant to Galactic Archaeology, illustrating its potential scientific value.

\section*{Acknowledgements}
KRS would like to acknowledge helpful discussions with Sabine Bolwell.    
We gratefully acknowledge support from the Australian Research Council through Laureate Fellowship FL220100117. D.H. acknowledges support from the National Aeronautics and Space Administration (80NSSC21K0652) and the Australian Research Council (FT200100871). D.S. is supported by the Australian Research Council (DP190100666). This work has made use of data from the European Space Agency (ESA) mission Gaia (\url{https://www.cosmos.esa.int/gaia}), processed by the Gaia Data Processing and Analysis Consortium (DPAC, \url{https://www.cosmos.esa.int/web/gaia/dpac/consortium}). Funding for the DPAC has been provided by national institutions, in particular the institutions participating in the Gaia Multilateral Agreement. This paper includes data collected with the TESS mission, obtained from the MAST data archive at the Space Telescope Science Institute (STScI). Funding for the TESS mission is provided by the NASA Explorer Program. STScI is operated by the Association of Universities for Research in Astronomy, Inc., under NASA contract NAS 5–26555. This work made use of several publicly available {\sc python} packages: {\tt astropy} \citep{astropy:2013,astropy:2018}, 
{\tt lightkurve} \citep{lightkurve2018},
{\tt matplotlib} \citep{matplotlib2007}, 
{\tt numpy} \citep{numpy2020}, and 
{\tt scipy} \citep{scipy2020}. 
This work has made an extensive use of
Topcat (\citealt{topcat}, \url{https://www.star.bristol.ac.uk/~mbt/topcat/}).

\section*{Data Availability}
The catalogue described in this article is available on the MNRAS website and CDS VizieR. Users should review the associated flags when utilizing the catalogue. The TESS data underlying this article are available at the MAST Portal (Barbara A. Mikulski Archive for Space Telescopes), at \url{https://mast.stsci.edu/portal/Mashup/Clients/Mast/Portal.html}.




\bibliographystyle{mnras}
\bibliography{sample} 








\bsp	
\label{lastpage}
\end{document}